\documentclass[11pt,onecolumn,superscriptaddress,notitlepage]{article}

\usepackage[total={6.5in,9in}, top=1.0in, includefoot]{geometry}
\usepackage{amsmath,amssymb}
\usepackage{hyperref}
\hypersetup{colorlinks,
    breaklinks,
    linkcolor=Red,
    urlcolor=Blue,
    anchorcolor=black,
    citecolor=Green,
    unicode=false}
\usepackage{graphicx}
\usepackage{wrapfig}
\usepackage{subcaption}
\usepackage{parskip}
\usepackage{cite}
\usepackage[dvipsnames,svgnames,table]{xcolor}
\usepackage{hyperref}
\usepackage{multirow}
\usepackage{times}
\usepackage[affil-it]{authblk} 
\usepackage{placeins}
\usepackage{colortbl}   
\usepackage{xcolor}     
\usepackage{booktabs}   
\usepackage{minted}
\usepackage{lineno}

\newcommand{\VE}{\textit{VE} }
\newcommand{\VEnospace}{\textit{VE}}
\newcommand{\VEmod}{$\textit{VE}_\text{modeled,R}$}
\newcommand{\VEmodspace}{$\textit{VE}_\text{modeled,R}\ $}
\newcommand{\VEmodmath}{\textit{VE}_\text{modeled,R}}
\newcommand{\VEmeas}{$\textit{VE}_\text{measured,R}$}
\newcommand{\VEmeasspace}{$\textit{VE}_\text{measured,R}\ $}
\newcommand{\VEmeasmath}{\textit{VE}_\text{measured,R}}
\newcommand{\VEmeashaz}{$\textit{VE}_\text{measured,H}$}

\newcommand{\VEmeashazmath}{\textit{VE}_\text{measured,H}}
\newcommand{\VEind}{$\alpha_\text{individual}$}
\newcommand{\VEindspace}{$\alpha_\text{individual}\ $}
\newcommand{\VEindmath}{\alpha_\text{individual}}
\newcommand{\alphaadj}{$\alpha_\text{adjusted}$}
\newcommand{\alphaadjspace}{$\alpha_\text{adjusted}\ $}
\newcommand{\alphaadjmath}{\alpha_\text{adjusted}}

\title{Incorporating vaccine effects into epidemiological models: common pitfalls and solutions}

\author[ ]{Casey E. Middleton}
\author[a,b]{Oliver Eales}
\author[a,b]{James M. McCaw}
\author[a,c]{Freya M. Shearer}

\affil[a]{Infectious Disease Dynamics Unit, Centre for Epidemiology and Biostatistics, Melbourne School of Population and Global Health, The University of Melbourne, Melbourne, Victoria, Australia}
\affil[b]{School of Mathematics and Statistics, The University of Melbourne, Melbourne, Victoria, Australia}
\affil[c]{Infectious Disease Ecology and Modelling, The Kids Research Institute, Perth, Western Australia, Australia }

\date{}

\begin{document}
\maketitle

\begin{abstract}
    Incorporating vaccination into mathematical models appears deceptively simple: models integrate vaccine-derived protections, such as reduced susceptibility to infection, using parameters informed by empirical estimates of vaccine efficacy or effectiveness (\VEnospace).
    In practice, however, empirical \VE estimates often do not correspond directly to the parameters of epidemiological models.
    Here, we extend previous work to demonstrate that in order to accurately parameterize a model, one must consider both a vaccine's mechanism of action and the statistic used to infer \VE from empirical data.
    When a vaccine confers leaky protection --- that is, vaccination partially rather than completely reduces individual infection risk --- we show that common empirical \VE estimation methods do not provide directly applicable values for model parameters.
    Na\"{\i}ve (i.e. direct) incorporation of these \VE estimates into models results in an underestimate of population-level vaccine impact.
    To make progress when these estimates are the only available sources for \VEnospace, we introduce a parameterization approach which more accurately aligns the modeled effect of vaccination with empirical estimates.
    Under this adjusted parameterization approach, models predict fewer total infections and lower herd immunity thresholds for leaky vaccines than would be predicted under current parameterization practices.
    Our parameterization guidelines and adjustment approach can be used to improve accuracy in  models that are used in vaccine decision making and public health planning.
\end{abstract}

\maketitle

\section{Introduction}
Vaccination is one of the most successful public health interventions to fight transmissible diseases~\cite{WHO2024Immunization}, and mathematical models are an integral tool used to guide vaccine decision making. Including the effects of vaccination in models of disease transmission empowers public health practitioners to weigh decisions on vaccine allocation policies~\cite{Bubar2021Model,Moore2021Modelling,Kisdi2024Optimal,Taira2004Evaluating,Elbasha2007Model,Anderson1983Vaccination}, predict herd immunity thresholds~\cite{Moore2021Vaccination,BubarMiddleton2022,Gay2004theory,Mossong2000Estimation}, and project expected future disease burden with (versus without) vaccination~\cite{Kiang2025Modeling,Weycker2005Population,Atchison2010Modelling,Lipsitch1997Vaccination,Sandmann2021Evaluating}. 

To incorporate vaccination into transmission models, an assumption must be made about the vaccine's mechanism of action. All-or-nothing vaccines provide perfect protection (i.e. sterilizing immunity) to some fraction of vaccine recipients, while the remaining fraction receive no protective benefit~\cite{SMITH1984}. Leaky vaccines provide some level of imperfect protection to all vaccinated individuals. In-between these two extremes exist so-called ``heterogeneous'' vaccines, where vaccine-derived protection varies among vaccine recipients~\cite{Gomes2014Missing,Nikas2023Competing,Langwig2017Vaccine,Halloran1992Interpretation}. 

For any assumed vaccine mechanism, developing a mathematical model relies on understanding the vaccine effect parameter(s), which specify the extent to which a vaccine reduces infection risk. Modelers face myriad potential pitfalls when parameterizing such models, because the effects of vaccination as measured by clinical trials and observational studies rarely correspond directly to the parameters of classic epidemiological models. While the existing literature demonstrates that \VE estimates may vary depending on both the statistical approach used to infer \VE from data and the vaccine mechanism of action~\cite{Halloran1997Study,SMITH1984,Nikas2023Competing,Shim2012Distinguishing}, it does little to guide researchers in best practices for parameterizing mathematical models of vaccination.

Here, we describe the \VE sources which should preferentially be used to parameterize vaccine effects in mathematical models of disease transmission, depending on a vaccine's assumed mechanism of action. We review common statistical methods for estimating \VEnospace, and demonstrate how na\"{\i}ve (i.e., direct) use of these estimates for parameterization of transmission models is incorrect for some vaccine mechanisms. We then introduce a parameterization approach to align model parameters with available \VE estimators. Using a simulation study, we demonstrate that our approach can be used to improve model accuracy in most scenarios and show that theoretical herd immunity thresholds are lower under this adjusted parameterization approach. Models that use this parameterization approach will be more accurate in their predictions for vaccine decision making and public health planning.


\section{Measuring vaccine effects}
Quantifying a vaccine's direct effect at preventing infection or disease is a fundamental need for public health decision making. During phase 3 of vaccine clinical trials, a first estimate is made of vaccine-derived protection in trial participants, called vaccine efficacy~\cite{Shrestha2007Safety,Polack2020Safety,Heath2021Safety}. Once vaccines are available to the general public, ``real-world" protection estimates are obtained from observational studies~\cite{Jackson2017Influenza,Bernal2021Effectiveness,Cohen2022BNT}. These estimates are called vaccine effectiveness. Both vaccine efficacy and effectiveness are measurements of vaccine-derived risk reduction, and in this manuscript we use \VE interchangeably to refer to either metric. Clinical trials and observational studies can measure \VE against a range of outcomes and often have a clinical endpoint. Here, we focus on \VE against infection acquisition  given its importance in transmission model dynamics.

In order to estimate \VE from data, it must be defined mathematically. The simplest way to define \VE is the reduction in cumulative infection risk for vaccinated individuals, as compared to unvaccinated individuals, measured using the cumulative attack rate ratio (ARR)~\cite{Halloran1997Study,Polack2020Safety,Shrestha2007Safety}. We refer to these cumulative risk-based estimates as \VEmeas. Other studies define \VE using hazards ratio (HR) estimators~\cite{Ohmit2013,Gram2022Vaccine,Shrestha2007Safety}, which we refer to as \VEmeashaz. HR estimates require data on transmission chains~\cite{Ohmit2013} or infection time~\cite{Gram2022Vaccine,Shrestha2007Safety} to calculate instantaneous risk, or hazard, conditioned on susceptibility. ARR estimates require less stringent monitoring and are thus the target estimated by most studies, including clinical trials~\cite{Shrestha2007Safety,Polack2020Safety,Heath2021Safety} and test-negative designs with no time component~\cite{Jackson2017Influenza,Lewnard2018Measurement}. These statistical definitions may yield different estimates of \VE depending on a vaccine's mechanism of action~\cite{SMITH1984,Nikas2023Competing}.

\subsection{Measured \VE depends on vaccine mechanism of action and study statistic}

We consider three mechanisms of action to describe the distribution of vaccine-derived protection amongst vaccine recipients: all-or-nothing, leaky, and heterogeneous. Smith, et. al. showed that ARR-based \VE estimates may differ from HR-based estimates for all-or-nothing and leaky vaccines~\cite{SMITH1984}. They find that for all-or-nothing vaccines, \VE estimates can accurately recover the direct effect of vaccination using ARR estimators, but will vary with time and converge to $\VEmeashazmath\to 1$ (indicating perfect protection) using HR estimators. For leaky vaccines, HR estimators can be used to recover the direct effect, but ARR estimators will converge to $\VEmeasmath\to 0$ (indicating zero protection) as study follow-up time increases, since an individual vaccinated with a leaky vaccine is guaranteed to be infected given infinite exposure opportunities. Thus, ARR estimators may--- due to differential depletion of susceptibles in vaccinated versus unvaccinated groups---incorrectly indicate waning of protection for a leaky vaccine through time, even when true protection is constant~\cite{Kahn2019Analyzing,Kahn2022Identifying,Goldstein2017Temporally,SMITH1984}. For heterogeneous vaccines, the ability of HR estimators to recover the direct effect has been shown to depend on the population-level distribution of vaccine-derived protection~\cite{Nikas2023Competing}.

To quantify the temporal variation in \VEmeasspace estimates, we developed a simple model of exposure events as a series of Bernoulli trials (see Supplemental Text S1.1 for details). From this model, we observe that \VEmeasspace decreases non-linearly with the number of exposure events for leaky and heterogeneous vaccines (Figs.~\ref{sfig:scatter_VE_bias},\ref{sfig:heatmap}) and does not necessarily approach zero if exposure events do not continue infinitely (Fig.~\ref{sfig:study_design}). These findings demonstrate the need for a deeper understanding of which \VE study statistics align with the parameters of disease transmission models (see Table~\ref{stab:param_sources}), to which we now turn.

\begin{table}[H]
\centering
\begin{tabular}{c|l|l}
\textbf{mechanism} & \multicolumn{1}{c|}{\textbf{direct incorporation}}                                            & \multicolumn{1}{c}{\textbf{adjustment needed}}                         \\ \hline
all-or-nothing                     & \begin{tabular}[c]{@{}l@{}} 
cumulative attack rate~\cite{SMITH1984} \end{tabular} & infection hazard~\cite{SMITH1984}                                                                   \\ \hline
heterogeneous                      &                                                       & \begin{tabular}[c]{@{}l@{}} 
I. infection hazard~\cite{Nikas2023Competing}  \\ 
II. cumulative attack rate$^*$  \end{tabular} \\ \hline
leaky                              & \begin{tabular}[c]{@{}l@{}} 
infection hazard~\cite{SMITH1984}   \end{tabular}                 & cumulative attack rate~\cite{SMITH1984}$^*$                  
\end{tabular}
\caption{Preferred study statistics for \VE to inform transmission model parameters. For each assumed vaccine mechanism, \VE sources may use a study statistic that directly corresponds to model parameters (direct incorporation), or that require model parameterization to be adjusted  for incorporation (adjustment needed). When multiple sources are possible, they are listed in order of preference. References that describe \VE biases for different study statistics and vaccine mechanism combinations are also indicated. $^*$ In Sections~\ref{sec:leaky_adj} and ~\ref{sec:hetero_adj}, we demonstrate an adjusted model parameterization approach when only cumulative attack rate-based estimators are available and a vaccine is believed to be leaky or heterogeneous, respectively.}
\label{stab:param_sources}
\end{table}


\section{Modeling vaccine effects: mismatch between empirical \VE estimates and model parameters}
We use the classic SIR model implemented as a system of ordinary differential equations to demonstrate the mismatch between empirical \VE estimates and model vaccine effect parameters~\cite{KeelingRohani2008}. The SIR model classifies each individual as either susceptible (S), infectious (I), or recovered (R). Susceptible individuals experience a time-varying infection hazard of $\lambda=\beta I/N$~\cite{Kenah2008Generation,Hens2010Seventy}, where $\beta$ is the transmission probability and $N$ is the total population size. Infectious individuals recover at rate $\gamma$. 

Additional model compartments are added for vaccinated individuals, with the exact model structure depending on a vaccine's mechanism of action (see Supplemental Text S1.2 for full model descriptions). For a given vaccine mechanism, we define the modeled population-level effect of vaccination (\VEmod$(t)$) as one minus the cumulative ARR in vaccinated versus unvaccinated groups at time $t$. For all vaccine mechanisms, infection is initially seeded in the unvaccinated population, so a brief burn-in period is observed for \VEmod$(t)$ as a modeling artefact while infections accumulate in the vaccinated population.

\begin{figure}
    \centering
    \includegraphics[width=0.75\linewidth]{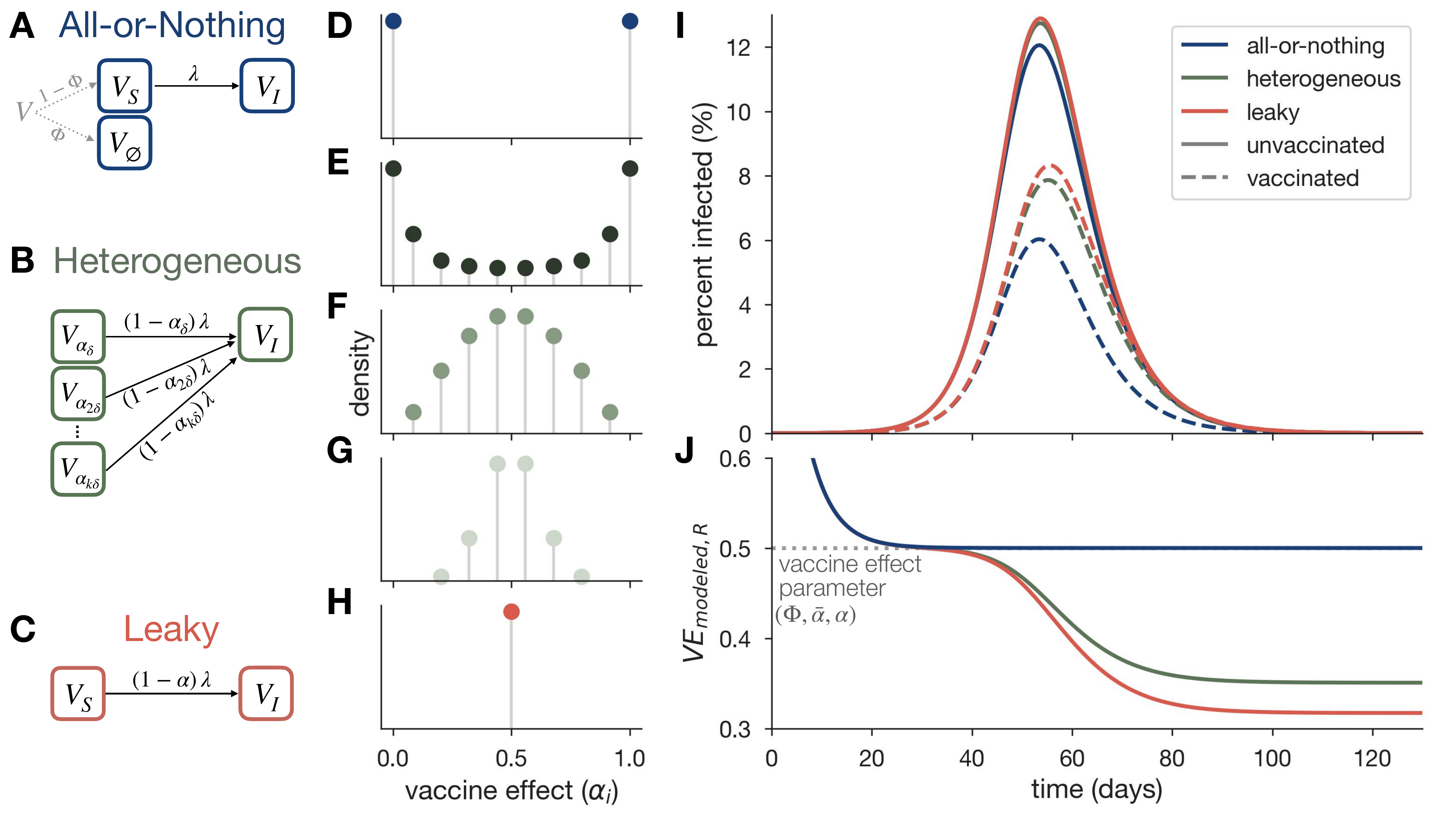}
    \caption{{\bf Infection dynamics vary based on vaccine mechanism of action.} Partial model diagrams (A--C) and vaccine-derived protection distributions (D--H; $c=10^{-4},\,0.4,\, 5,\, 20,\, 10^4$) for all-or-nothing, heterogeneous, and leaky vaccine mechanisms. Temporal infection dynamics for vaccinated (dashed) and unvaccinated (solid) populations (I) and time-varying \VEmodspace (J) for each vaccine mechanism are depicted under the same parameterization ($\Phi=\bar{\alpha}=\alpha=0.5$, $R_0=2.2$, 35\% vaccinated at start of simulation).}
    \label{fig:time_varying_VE}
\end{figure}

\subsection{Modeling vaccination}
All-or-nothing vaccine models assume that vaccination provides perfect protection against infection for some proportion ($\Phi$) of the vaccinated population, and none to the remaining proportion ($1-\Phi$). This results in a split vaccinated population; some are fully protected and do not participate in transmission dynamics ($V_{\emptyset}$), while others remain fully susceptible ($V_S$) and thus participate identically to unvaccinated susceptible individuals (Fig.~\ref{fig:time_varying_VE}A,D). Infection dynamics for vaccinated individuals are a scaled version of the unvaccinated dynamics, reduced by a factor $(1-\Phi)$, with identical peak times (Fig.~\ref{fig:time_varying_VE}I). For all-or-nothing vaccines, we observe a time-invariant $\VEmodmath=\Phi$ (Fig.~\ref{fig:time_varying_VE}J).

Leaky vaccine models assume that vaccination reduces infection hazard by a factor of $\alpha$ for all vaccinated individuals\ (Fig.~\ref{fig:time_varying_VE}C,H). The peak of vaccinated infections occurs slightly later than the peak of unvaccinated infections (Fig.~\ref{fig:time_varying_VE}I), and \VEmodspace decreases over the course of the epidemic (Fig.~\ref{fig:time_varying_VE}J). Early in an outbreak, $\VEmodmath=\alpha$. As the epidemic progresses, we observe an S-shaped decline in \VEmod, such that $\VEmodmath<\alpha$, mirroring the results of Smith, Rodrigues, and Fine~\cite{SMITH1984}. When compared to all-or-nothing vaccines, leaky vaccines result in more infections amongst the vaccinated population under the same vaccine-derived protection parameter ($\alpha = \Phi$, Fig.~\ref{fig:time_varying_VE}I)~\cite{SMITH1984,Halloran1997Study,Lee2025Vaccine}.

Heterogeneous vaccine models are more general and assume that vaccine-derived protection varies among vaccine recipients. Mathematically, the effect of heterogeneous vaccines may be described as individual-level hazard reductions $\alpha_i$, defined by the probability distribution $P(\alpha_i)$. Here, we consider protection to be beta distributed across vaccine recipients, $P(\alpha_i) \sim \text{Beta}(\bar{\alpha},c)$, where $\bar{\alpha}$ is the mean hazard reduction provided by a vaccine, and $c>0$ determines the shape of the distribution~\cite{Gomes2014Missing,Langwig2017Vaccine}. The beta distribution has flexibility to model all-or-nothing ($c\to0$, $P(\alpha)>0$ at $\alpha=0,1$; Fig.~\ref{fig:time_varying_VE}D) and leaky ($c\to\infty$, $P(\alpha)>0$ at $\alpha=\bar{\alpha}$; Fig.~\ref{fig:time_varying_VE}H) vaccine-derived protection distributions. When $c\in(0,\infty)$, individual vaccine-derived protection lies between all-or-nothing and leaky. For example, when $\bar{\alpha}=0.5$ and $c=10$ (Fig.~\ref{fig:time_varying_VE}F), the heterogeneous model exhibits similar dynamics to the leaky model when $\alpha = 0.5$, though fewer total infections occur in vaccinated individuals (Fig.~\ref{fig:time_varying_VE}I). Like the leaky model, \VEmodspace appears to wane over time, though the final \VEmodspace is slightly higher than for the leaky model (Fig.~\ref{fig:time_varying_VE}J). These dynamics vary depending on the shape parameter $c$ (Fig.~\ref{sfig:time-varying-VE-supp}).

\subsection{Identifying the mismatch between empirical \VE estimates and model parameters}
Regardless of a vaccine's mechanism of action, models of vaccination rely on \VE estimates to inform their respective vaccine effect parameters. The most common \VE estimators are ARR-based, which is an accurate informant for all-or-nothing vaccine models (Table~\ref{stab:param_sources}). However, leaky and heterogeneous vaccine effect parameters act as scaling constants which reduce the relative infection hazard of vaccinated individuals. Ideally, these parameters should be informed using a study design which estimates per-exposure or instantaneous risk reduction, such as a challenge study or HR estimator~\cite{Halloran1997Study,Rhodes1996Counting}. While these estimators can be used to directly inform leaky model parameters, they may exhibit varying degrees of bias for heterogeneous models depending on the distribution of vaccine-derived protection (i.e., parameter $c$)~\cite{Nikas2023Competing}.

\begin{figure}[h!]
    \centering
    \includegraphics[width=0.8\linewidth]{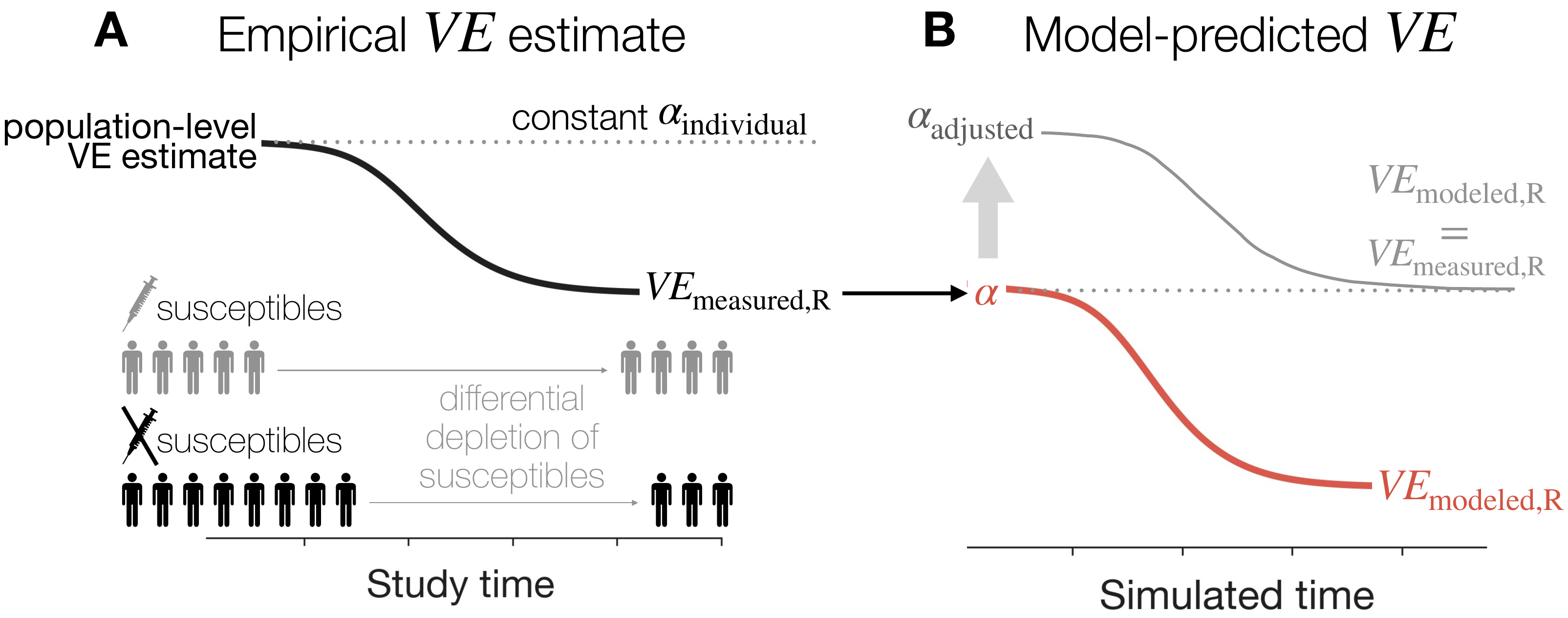}
    \caption{{\bf The pipeline of empirical to modeled vaccine effectiveness (\VEnospace) estimates for a leaky vaccine.} (A) Empirical \VE estimates using risk-based (ARR-based) estimators are prone to waning \VEmeasspace over time, despite a constant true protection level (\VEind), due to differential depletion of susceptibles in vaccinated and unvaccinated populations. (B) When \VEmeasspace is used to inform a leaky model's vaccine effect parameter ($\alpha$), depletion of susceptible bias occurs again in the modeled population. Adjusting parameterization (\alphaadj, see Section~\ref{sec:leaky_adj}) aligns \VEmodspace with \VEmeas.}
    \label{fig:schematic}
\end{figure}

In many cases modelers may only have access to risk-based (ARR-based) estimates of \VEmeas. If a vaccine is believed to be leaky with a constant hazard reduction of \VEindspace and study participants experience multiple exposure events, \VEmeasspace estimates will be lower than \VEindspace (Fig.~\ref{fig:schematic}A). Thus, directly incorporating \VEmeasspace as a model's vaccine effect parameter ($\alpha$, Fig.~\ref{fig:schematic}B), as is common practice, sets a false equivalence between the cumulative risk ratio \VEmeasspace and the hazard ratio $\alpha$. The modeled population-level vaccine effect \VEmodspace is likely to underestimate the true population-level impact of vaccination using this approach. The same is true for heterogeneous vaccines. Thus, when only ARR-based estimators are available, modelers should seek to adjust their model parameterization accordingly.



\section{Adjusted model parameterization approach}
We demonstrate an approach to adjust model parameterization when only ARR-based estimators are available and a vaccine is believed to be leaky or heterogeneous. This approach operates under the assumption that if we only have access to a risk-based (ARR-based) estimator \VEmeas, we should align our modeled population-level vaccine effect (\VEmod) with this estimator (Fig.~\ref{fig:schematic}B, gray curve).

\subsection{Adjusted leaky model parameterization}
\label{sec:leaky_adj}
Shim and Galvani introduce an adjusted parameterization approach to align \VEmodspace and \VEmeasspace in leaky models, which we describe here~\cite{Shim2012Distinguishing}. Given that \VEmodspace is defined using the cumulative attack rate ratio, we begin by solving for the modeled cumulative attack rates (i.e., final size) in both unvaccinated ($\Omega_U$) and vaccinated ($\Omega_V$) populations. For the leaky SIR model presented here, these quantities are defined by the following transcendental equations:
\begin{equation}
\begin{aligned}
\Omega_U &= 1 - e^{-R_0\, ((1-\nu)\, \Omega_U + \nu\, \Omega_V)}, \\[6pt]
\Omega_V &= 1 - e^{-(1-\alpha)\,R_0\, ((1-\nu)\, \Omega_U + \nu\, \Omega_V)},
\label{eq:final_size}
\end{aligned}
\end{equation}
where $\nu$ is the vaccinated fraction of the population and $R_0$ is the basic reproduction number. \VEmodspace is defined as $\VEmodmath = 1 - \Omega_V / \Omega_U$.

We then use a root finder approach to solve for the value of \alphaadjspace which most closely satisfies \VEmod = \VEmeasspace for a given parameter set (see Supplemental Text S1.3.1 for details). The resulting $\alphaadjmath > \alpha$ (Fig.~\ref{fig:schematic}B, gray arrow) produces a modeled population-level \VEmodspace that aligns with the real-world population-level estimate of \VEmeas. 

To demonstrate the approach, we consider a representative scenario where $R_0=2.6$, $\VEmeasmath=0.3$, and the population has 60\% vaccine coverage. The leaky SIR model under standard parameterization ($\alpha = \VEmeasmath$) predicts that 82\% of the population will be infected in this scenario. By applying our adjusted parameterization approach, we find that $\alphaadjmath = 0.51$, which results in a slower epidemic growth rate and lower epidemic peak, such that only 68\% of the population is infected by the end of the modeled outbreak.

\subsection{Adjusted heterogeneous model parameterization}
\label{sec:hetero_adj}
The heterogeneous model describes variable vaccine-derived protection against infection for vaccinated individuals and has a similar compounding depletion of susceptible bias as the leaky model. Here, we extend the methodology presented by Shim and Galvani for the leaky model~\cite{Shim2012Distinguishing} to parameterize vaccine effect in a heterogeneous modeling framework. Empirical estimates of \VEmeasspace typically represent a population-level average (and uncertainty around the estimate of an average) for vaccine-derived protection --- that is, they do not capture individual-level heterogeneity in protection. We have limited understanding of the true distribution of vaccine-derived protection across vaccinated individuals because it is difficult to measure~\cite{Gomes2014Missing,Lee2025Vaccine}. Therefore, we focus here on aligning average \VEmeasspace with \VEmod, assuming protection is beta distributed with a given shape parameter $c$.

For the heterogeneous model, the cumulative attack rate for vaccinated populations is defined as 
$$ \omega_V = \sum^i P(\alpha_i) \cdot \omega_{V_i}, $$
where $P(\alpha_i)$ is the proportion of the vaccinated population in compartment $i$. The quantity $\omega_{V_i}$ is the proportion of compartment $i$ that has been infected, i.e., the cumulative attack rate for compartment $i$, given by
$$ \omega_{V_i} = 1 - e^{-(1-\alpha_i)\, R_0\, ((1-\nu)\, \omega_U + \nu\, \omega_V)} $$ 
for the compartment-specific hazard reduction level $\alpha_i$. The cumulative attack rates for all unvaccinated and vaccinated populations are thus defined by
\begin{equation}
\begin{aligned}
\omega_U &= 1 - e^{-R_0 ((1-\nu) \omega_U + \nu\, \omega_V)}, \\[6pt]
\omega_V &= 1 - \sum^i P(\alpha_i) \cdot e^{-(1-\alpha_i)\, R_0\, ((1-\nu)\, \omega_U + \nu\, \omega_V)}.
\label{eq:final_size_frailty}
\end{aligned}
\end{equation}

For a given shape parameter $c$, we solve for the value of $\bar{\alpha}_{adjusted}$ which satisfies
\begin{equation}
    \VEmeasmath = \VEmodmath = 1-\omega_V/\omega_U
    \label{eq:frailty_relationship}
\end{equation}
using a root-finding approach (see Supplemental Text S1.3.2 for details).

\begin{figure}
    \centering
    \includegraphics[width=0.8\linewidth]{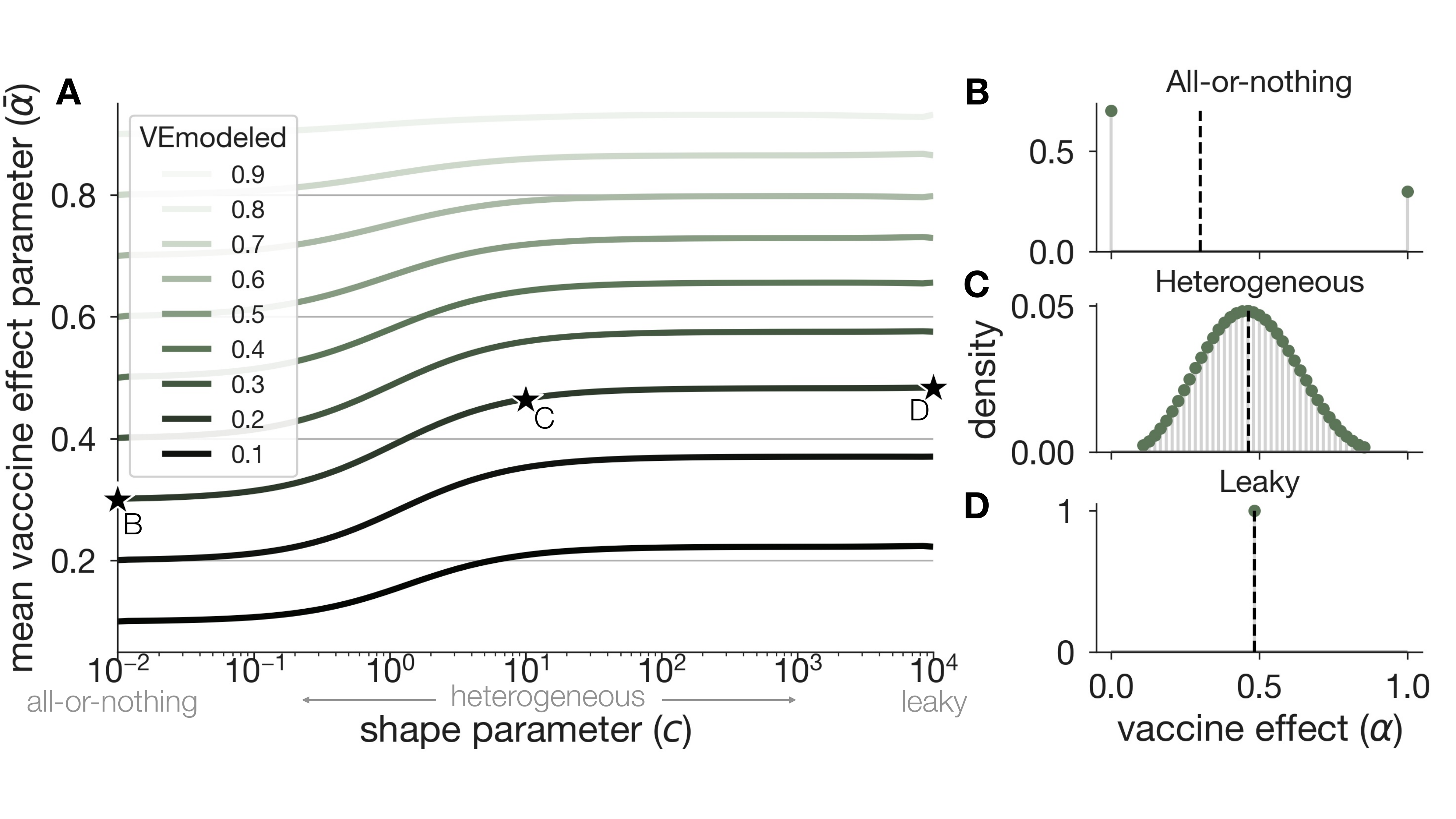}
    \caption{{\bf \VEmodspace depends on the distribution of vaccine-derived protection.} (A) Isoclines for \VEmodspace under a given mean vaccine effect parameter ($\bar{\alpha}$) and varying distribution shapes ($c$). Three sample distributions are shown that produce $\VEmodmath=0.3$ when $c\to 0$ (all-or-nothing, B), $c=10$ (heterogeneous, C), and $c\to\infty$ (leaky, D), with mean vaccine effect ($\bar{\alpha}$) depicted as a vertical dashed line. }
    \label{fig:frailty_isoclines}
\end{figure}

The value of $\bar{\alpha}_{adjusted}$ varies depending on the shape parameter $c$ (Fig.~\ref{fig:frailty_isoclines}A). Consider the scenario where $\VEmodmath=0.3$, all other model parameters are equal, and we do not know $c$. All-or-nothing vaccines ($c \to 0$) do not experience depletion of susceptible bias, so $\VEmodmath = \bar{\alpha} = 0.3$ (Fig.~\ref{fig:frailty_isoclines}B). Increasing the shape parameter to $c=10$ results in a heterogeneous model, for which a value of $\bar{\alpha}_{adjusted} = 0.46$ is needed to recover $\VEmodmath = 0.3$ (Fig.~\ref{fig:frailty_isoclines}C). Similarly, a leaky vaccine ($c\to\infty$) requires $\bar{\alpha}_{adjusted} = 0.48$ to recover $\VEmodmath = 0.3$ (Fig.~\ref{fig:frailty_isoclines}D).

The difference between \alphaadjspace and \VEmeasspace also depends on the transmission rate, vaccine coverage, the distribution of protection, and the value of \VEmeas. The greatest difference between \VEmeasspace and $\bar{\alpha}$ exists when $\bar{\alpha}= 0.48$ and when a vaccine has a leaky mechanism of action (Fig.~\ref{sfig:frailty_heatmap}). To test whether the adjusted parameterization aligns modeled and real-world vaccine effects, we turn to a simulation study.


\section{Evaluation of parameterization approach using simulated data}
For any given modelling study, the ideal parameterization would align the modeled vaccine effect distribution $P(\alpha)$ with the true underlying protection distribution. For example, a leaky vaccine model's $\alpha$ should align with \VEind, but we rarely have data measuring \VEindspace directly. Solving for \alphaadjspace using the approach presented here can approximate \VEindspace if the model assumptions closely align to the epidemiological context in which \VEmeasspace was studied (i.e., vaccine coverage, basic reproduction number, etc within the population at the time that the \VE study was conducted). Given that the magnitude of bias (i.e., the difference between \VEindspace and \VEmeas) depends on this epidemiological context (Fig.~\ref{sfig:heatmap}), \alphaadjspace may not align with the true \VEindspace if the model does not reflect this context. In a scenario where the epidemiological context of the \VE study does not match the model assumptions, does the adjusted parameterization improve accuracy in predicting the population-level impact of vaccination? We use simulated data to evaluate this question under a range of epidemiological scenarios. 

\subsection{Simulated study design}
We begin by generating synthetic experimental data to mirror empirical \VEmeasspace estimates. The data is generated using the SIR modeling framework described above for either leaky or heterogeneous vaccine mechanisms. We specify the vaccine coverage at model initialization ($v$), vaccine-derived hazard reduction (\VEind), basic reproduction number ($R_0$), and other known parameters. If the vaccine mechanism is heterogeneous, we also specify the shape parameter ($c$) and set $\bar{\alpha}=\VEindmath$. We simulate an epidemic under these parameters (see Supplemental Text S1.2 for details) and calculate \VEmeasspace using the cumulative attack rate ratio (ARR) in vaccinated versus unvaccinated populations. 

Next, we consider the result of using \VEmeasspace to parameterize a model of the same form. To do so, we specify a model parameter set consisting of: vaccine coverage ($v_\text{modeled}$), basic reproduction number ($R_{0,\text{modeled}}$), and shape parameter ($c_\text{modeled}$), if relevant. The modeled population-level vaccine effectiveness (\VEmod) is calculated using cumulative attack rates under three scenarios. First, $\textit{VE}_\text{true,R}$ is the modeled population-level vaccine effect if \VEindspace is used to inform the model's vaccine effect parameter. Second, $\textit{VE}_\text{standard,R}$ is the result of directly incorporating \VEmeasspace as the model's vaccine effect parameter, as is standard practice in the epidemiological modeling literature at the time of writing. Third, $\textit{VE}_\text{adjusted,R}$ is the result of applying the parameter adjustment approach described in Section 4.

The residual error is calculated as the difference between $\textit{VE}_\text{standard,R}$ or $\textit{VE}_\text{adjusted,R}$ and $\textit{VE}_\text{true,R}$, which quantifies how well a model parameterization captures the true population-level impact of vaccination. If small residual errors are observed, this would indicate that using \VEmeas, either directly or with our adjusted parameterization, to inform model parameters produces similar results to using \VEind, the ideal informant. If the adjusted parameterization improves model accuracy, we expect to observe a lower residual error under adjusted than standard parameterization.

\begin{figure}[h!]
    \centering
    \includegraphics[width=0.8\linewidth]{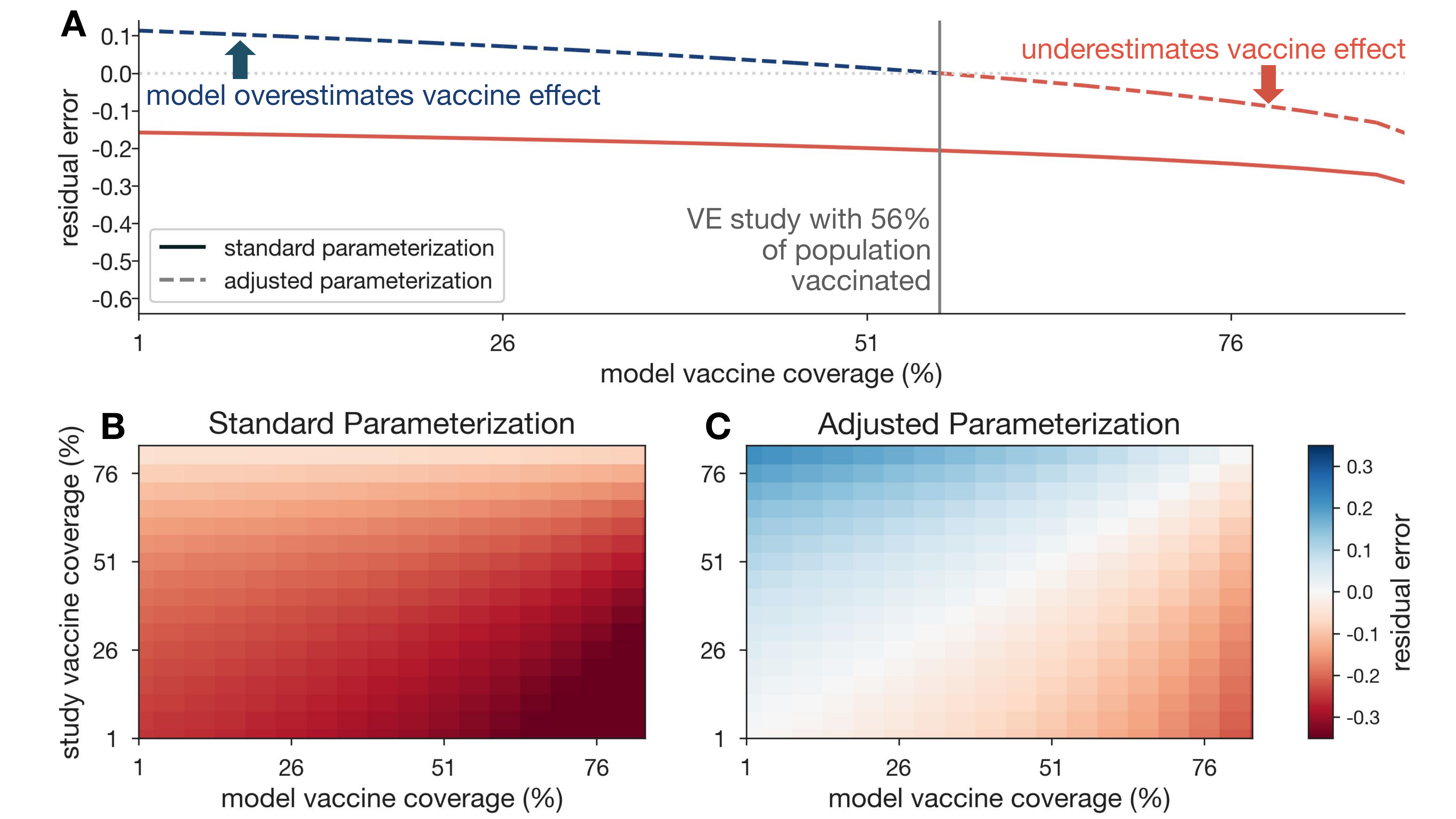}
    \caption{{\bf Model accuracy is high using the adjusted parameterization approach if model parameters align with study scenario.} (A) Residual error using a \VE estimate from a study setting with 56\% of the population vaccinated (vertical line). Error is depicted for standard parameterization ($\alpha = \VEmeasmath$, solid curve) and the adjusted parameterization (dashed curve) under a given model vaccine coverage. (B,C) Residual error as both study and model vaccine coverage vary, using standard (B) and adjusted (C) methods to parameterize $\alpha$. $R_0 = 2.6$ in all scenarios.}
    \label{fig:leaky_residual}
\end{figure}

\subsection{Evaluation of adjusted parameterization for leaky model}
The adjusted parameterization consistently recovers the true population-level vaccine effect when the precise epidemiological context of the \VEmeasspace estimation study is known, and reduces absolute residual error for almost all scenarios where it is not known. To demonstrate this, we analyze a scenario where vaccine coverage throughout the \VEmeasspace study is 56\% ($v=0.56$). The adjusted parameterization recovers the true effect of vaccination (i.e., the residual is zero) when \alphaadjspace is computed under a model scenario with the same $R_0$ and vaccine coverage ($v_\text{modeled}=v=0.56$, Fig.~\ref{fig:leaky_residual}A, vertical line). If the model scenario does not align with the study's vaccine coverage, the adjusted parameterization may lead to an overestimate (when $v_\text{modeled}<v$) or underestimate (when $v_\text{modeled}>v$) of the true vaccine effect. However, even when the vaccine coverage is not matched, the adjusted parameterization still reduces absolute residual error when compared to standard parameterization~(Fig.~\ref{fig:leaky_residual}A).

By varying study vaccine coverage, we observe that standard parameterization is most biased when $v$ is low, such as when a model relies on clinical trial \VEmeasspace estimates (Fig.~\ref{fig:leaky_residual}B). The adjusted parameterization almost always decreases absolute residual error but may increase error when study vaccine coverage is high and \alphaadjspace is computed under a model scenario with low vaccine coverage (Fig.~\ref{fig:leaky_residual}B,C). If other model parameters, such as $R_0$, are misaligned with their value at the time of the \VEmeasspace study, the adjusted parameterization may not achieve a residual error of zero but generally still reduces absolute residual error (Fig.~\ref{sfig:correction_accuracy}). 

These results demonstrate lower residual error when the modeled scenario is closely aligned to the study scenario, suggesting that modelers solve for \alphaadjspace using a single parameter set they believe is most closely aligned to the epidemiological scenario at the time \VEmeasspace was studied. This value of \alphaadjspace can then be used to simulate different epidemiological scenarios to improve model prediction accuracy.

\subsection{Evaluation of adjusted parameterization for heterogeneous model}
Uncertainty around the distribution of protection can reduce the accuracy of the adjusted parameterization. Since empirical studies do not capture heterogeneity in protection, the distribution of protection (here, shape parameter $c$) is a large source of uncertainty for a heterogeneous model~\cite{Gomes2014Missing}. All-or-nothing and leaky vaccine models are implicitly assuming a particular distribution of vaccine-derived protection. To test how this uncertainty may impact the accuracy of the adjusted parameterization, we simulate data for a heterogeneous vaccine using a given parameter set ($R_0 = 2.6$, $v=0.35$, $\bar{\alpha}=0.7$) and vary the vaccine-derived protection shape parameter ($c_{true}$) to simulate outbreak data and calculate \VEmeas. We then use \VEmeasspace to inform the average vaccine effect parameter $\bar{\alpha}$ of the heterogeneous model under our adjusted parameterization, while varying the shape parameter ($c_{model}$) with otherwise identical parameter conditions.

\begin{figure}[h!]
    \centering
\includegraphics[width=0.65\linewidth]{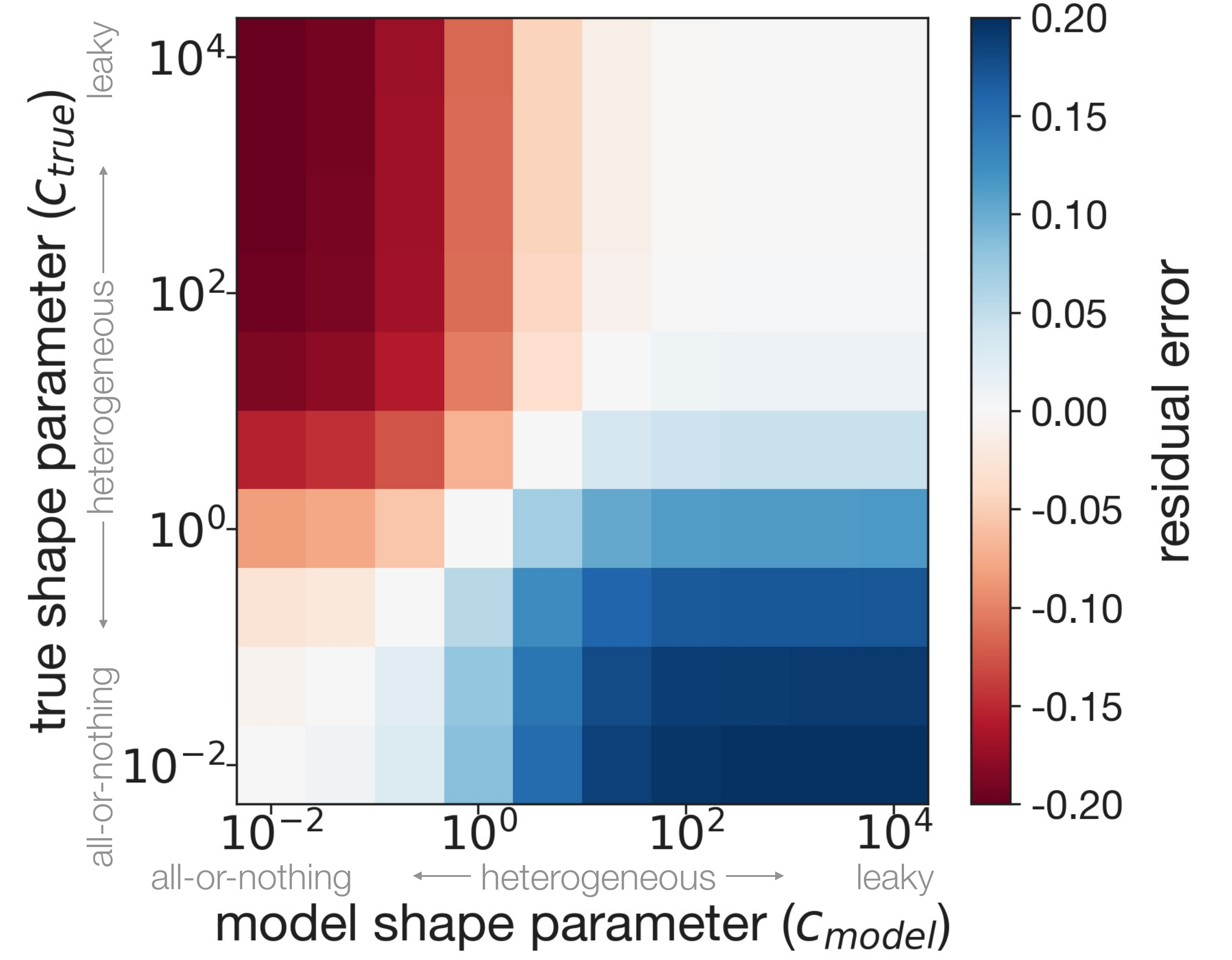}
    \caption{{\bf Adjusted parameterization accuracy is high for a heterogeneous vaccine if model correctly assumes true distribution shape.} Residual error from the adjusted parameterization is shown, using a \VE estimate from a study setting with a given shape parameter ($c_{true}$). The study assumes $R_0=2.6$ with a 35\% vaccinated population and an average vaccine effect of $\bar{\alpha}=0.7$. The model assumes an identical $R_0$ and vaccinated fraction, with varying $c_{model}$.}
    \label{fig:heterogeneous_residual}
\end{figure}

The adjusted parameterization has a residual of zero when the model correctly assumes the distribution shape ($c_{model} = c_{true}$, Fig.~\ref{fig:heterogeneous_residual}, diagonal). However, residual error may increase if there are inconsistencies between $c_{true}$ and $c_{model}$. The greatest residual error occurs when the true distribution is either leaky or all-or-nothing, and the model assumes the opposite. Uncertainty may also exist in other parameters, such as $R_0$, which can change the magnitude of residual error and reduce the parameter space in which the adjusted parameterization approach achieves a residual error of zero (Fig.~\ref{sfig:heterogeneous_correction_heatmaps}). The adjusted parameterization reduces absolute residual error for the heterogeneous model if the assumed distribution of protection is closely aligned to its true distribution but may increase residual error if this assumed distribution is not aligned with its true shape.


\section{The impact of incorrect parameterization on herd immunity thresholds}
We have shown that adopting the adjusted parameterization approach presented here will reduce a model's residual error, but does this adjustment notably change model predictions? We use herd immunity threshold (HIT) estimation as a case study to investigate this question. For a leaky vaccine under SIR model assumptions, the HIT is given by $HIT = \frac{1}{\alpha} \left( 1 - \frac{1}{R_0} \right)$,
where $\alpha$ and $R_0$ are the model vaccine effect parameter and basic reproduction number, respectively. We calculate this threshold for three disease scenarios, including a range of basic reproduction numbers ($R_0=$ 14, 2, and 5) and cumulative risk-based \VEmeasspace estimates ($\VEmeasmath=$ 0.75, 0.5, and 0.85). These are plausible parameter scenarios for pertussis, pandemic influenza, and alpha-era COVID-19, respectively~\cite{Zhang2012, Wearing2009, Biggerstaff2014, Osterholm2012, LiuRocklov2022, Lin2025, LinkGellesetal2025}. However, we note that this analysis relies on simple SIR modeling assumptions and is included to demonstrate how the adjusted parameterization changes model predictions, not to establish real-world vaccination targets for these pathogens.

\begin{figure}[h!]
    \centering
\includegraphics[width=0.8\linewidth]{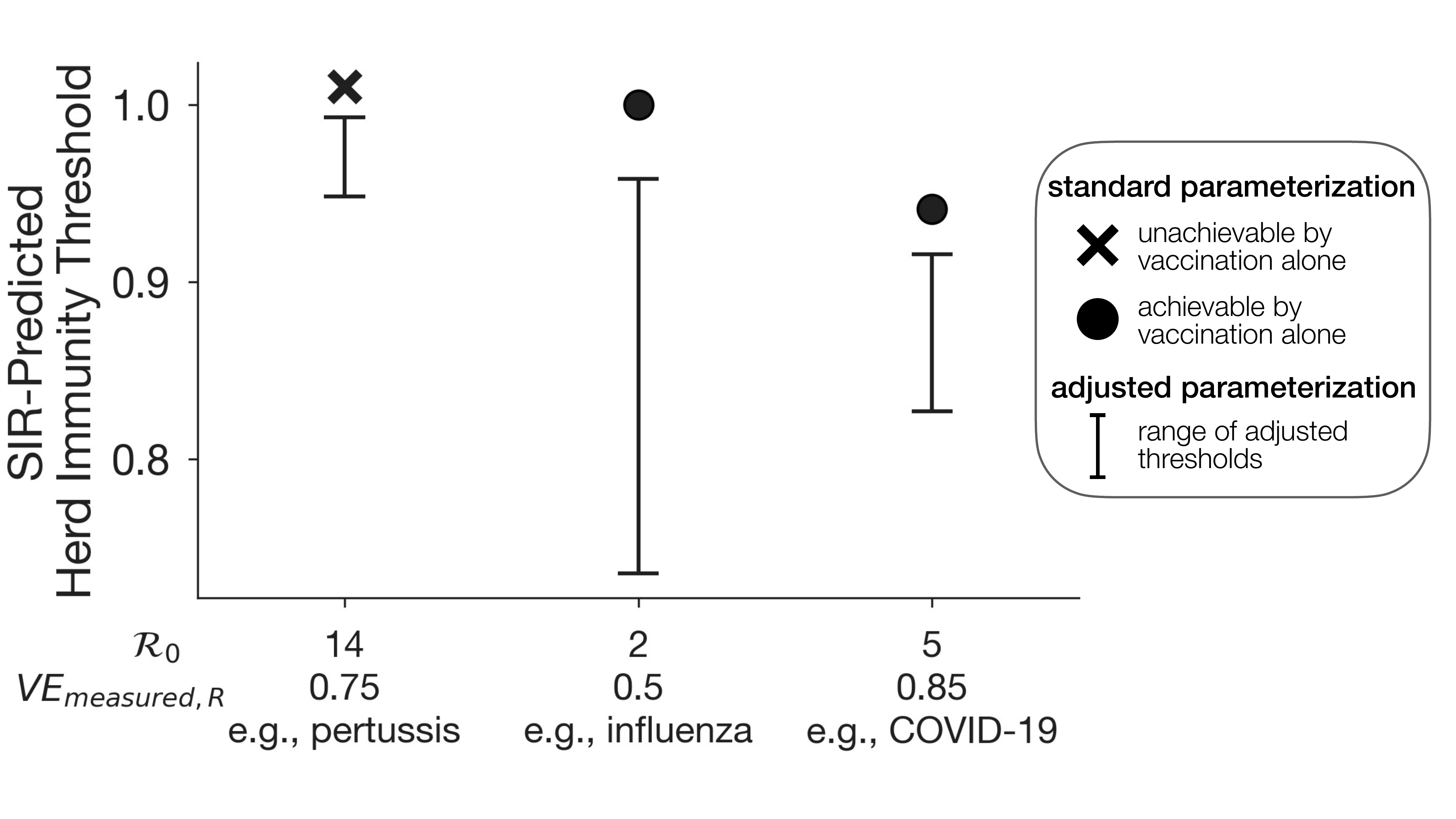}
    \caption{{\bf The adjusted parameterization leads to lower predicted herd immunity thresholds for a leaky model.} Predicted herd immunity threshold using the SIR model under standard parameterization (filled point or X) and adjusted parameterization range (lines and whiskers) for three $R_0$ and \VEmeasspace scenarios. Adjusted parameterization ranges are shown for \VE study vaccination coverage $v \in [0.01,0.99]$.}
    \label{fig:HIT}
\end{figure}

We first calculate the HIT using the standard parameterization method ($\alpha = \VEmeasmath$). The resulting thresholds indicate that herd immunity cannot be achieved by vaccination alone for the pertussis scenario (Fig.~\ref{fig:HIT}). The pandemic influenza scenario has a high but still theoretically achievable threshold, while COVID-19 is slightly lower. Next, we compute \alphaadjspace under a range of plausible study vaccine coverages between 0.01 and 0.95 to generate a range of adjusted HIT estimates. The resulting range of HIT estimates are consistently less than or equal to those estimated by the standard parameterization approach (Fig.~\ref{fig:HIT}). For the pertussis scenario, the adjusted parameterization results in a HIT that is achievable by vaccination alone. For the pandemic influenza and COVID-19 scenarios, adjusted parameterization reduces the vaccine coverage required to achieve herd immunity. The difference between standard and adjusted HITs and the variance amongst adjusted HIT estimates depends on both \VEmeasspace and $R_0$ (Supp. Fig~\ref{sfig:HIT_heatmaps}), since these parameters both affect the bias between \alphaadjspace and \VEmeasspace at a given vaccine coverage.


\section{Discussion}
We have demonstrated that when parameterizing mathematical models of vaccination, researchers must consider the statistic used to estimate \VE and the assumed vaccine mechanism of action (Table~\ref{stab:param_sources}). Models of all-or-nothing vaccines should preferentially use ARR estimators, while models of leaky vaccines should give preference to HR estimators of \VEnospace. Models of heterogeneous vaccines cannot be fully described under current \VE measurement approaches, but the average protection level $\bar{\alpha}$ is generally best parameterized using HR estimators. Given that cumulative risk-based (ARR-based) estimates are often the only available \VE source, we demonstrate an approach to parameterize leaky and heterogeneous models using risk-based \VEmeasspace to more accurately align modeled vaccine effects with measured \VE values. Models parameterized using this approach predict fewer total infections and lower herd immunity thresholds for a given \VEmeasspace relative to the standard parameterization approach, demonstrating the potential implications of parameterization approach on policy decision-making.

Our work builds on previous literature which established a hierarchy of data needs to inform direct protection from vaccination~\cite{Halloran1997Study}. Exposure controlled studies, such as challenge studies or contact tracing studies, are useful to distinguish exposure from susceptibility but are extremely costly to perform at scale~\cite{Halloran1997Study}. Cumulative incidence is the most straightforward data to collect~\cite{Halloran1997Study} but \VE estimates from these data appear to wane with time for leaky~\cite{SMITH1984,Lee2025Vaccine,Ragonnet2015Vaccination} or heterogeneous vaccines, despite true protection remaining constant. Findings that test-negative study designs cannot recover \VE for leaky vaccines are based on the same phenomenon~\cite{Lewnard2018Measurement}. HR-based estimators can help distinguish exposure from susceptibility~\cite{Halloran1997Study} but may appear to increase over the course of an epidemic for all-or-nothing~\cite{SMITH1984} and heterogeneous~\cite{Nikas2023Competing} vaccines. 

Despite previous work identifying cumulative risk-based estimators as poor informants for leaky model parameters~\cite{Shim2012Distinguishing,Haber1995Effect}, modelers must rely on these estimators when they are all that is available. Many published models used to inform vaccine allocation (e.g.,~\cite{Bubar2021Model,Weycker2005Population,Sandmann2021Evaluating}), dosing (e.g.,~\cite{Wood2009Optimal}), and herd immunity thresholds (e.g.~\cite{Moore2021Vaccination,BubarMiddleton2022}) have relied on risk-based \VEmeasspace estimates to inform leaky model parameters without adjustment. Our results show that this direct incorporation of \VEmeasspace to inform leaky model parameters may underestimate population-level vaccine effect. We have clarified the statistical methods which can reliably estimate these parameters and formalized the approach proposed by Shim and Galvani to adjust model parameterization when only risk-based estimators are available~\cite{Shim2012Distinguishing,UTAustinMeasles2026}. We have shown that this approach improves model accuracy and demonstrated that its incorporation will change model predictions of epidemic size and herd immunity thresholds.


Studies of \VEmeasspace are subject to a number of biases, which have been well documented elsewhere~\cite{Halloran1997Study,Halloran1992Interpretation,SMITH1984,Lewnard2018Measurement,Edlefsen2014leaky}. We focus our work on depletion of susceptible bias as the main contributor to the misalignment between \VEmeasspace estimates and model parameters. We also note that empirical studies used to estimate \VEmeasspace often focus on symptomatic (or ``medically-attended") illness --- these studies are inherently limited in their ability to inform vaccine-derived protection against infection acquisition. Furthermore, these studies are not designed to measure individual-level variation in response to vaccination. Our results, along with previous studies~\cite{Lee2025Vaccine,Nikas2023Competing,Shim2012Distinguishing}, demonstrate that understanding vaccine mechanism or the distribution of protection can greatly impact model results. All-or-nothing or leaky vaccines are common simplifying assumptions, which may not be valid representations of the complex individual-level immune response to vaccination. Novel \VE study designs and statistical methods that target individual-level protection would be valuable to the field. In the interim, model-based studies should make appropriate adjustments to parameterization and include sensitivity analyses to vaccine mechanism.

We have highlighted the presence of biases when parameterization is incorrect for even the simplest models of vaccination. More complex model designs, such as models including multi-factor or waning vaccine-derived protection, age structure, and stochasticity are expected to be prone to similar biases. We anticipate that these models require similar adjustments to parameterization when vaccines are assumed to provide leaky or heterogeneous protection and cumulative risk-based \VE estimators are used. Similarly, future work would benefit from the development of approaches to adjusted parameterization for all-or-nothing and heterogeneous models when HR estimators are the only available source of \VEnospace. 

Our work demonstrates that current standard parameterization practices for incorporating the effect of vaccination into transmission models may not accurately align model outcomes with empirical estimates of vaccine efficacy or effectiveness. Through this work, we have made progress toward increasing the accuracy of transmission models that incorporate vaccine effects, which will provide better evidence for public health decision making on vaccination.

\section*{Data Availability}
All open-source code (Python 3.9.7) needed to evaluate the conclusions in the paper is available at\\ \href{https://github.com/CaseyMiddleton/ModelingVE}{https://github.com/CaseyMiddleton/ModelingVE}. All data needed to evaluate the conclusions in the paper are present in the paper and/or the Supplementary Materials.

\section*{Acknowledgements}
We thank Dr Stephen Kissler and Alexander Pillai for their feedback on this manuscript, as well as the funders who made this work possible. Casey Middleton and James McCaw were supported by an ARC Laureate Fellowship (FL240100126). Oliver Eales was supported by a University of Melbourne McKenzie Fellowship. Freya Shearer was supported by an NHMRC Investigator Grant (Emerging Leader Fellowship, 2021/GNT2010051).

\newpage
\bibliographystyle{unsrt}
\bibliography{references}

\clearpage
\renewcommand{\thetable}{S\arabic{table}} 
\setcounter{table}{0}
\renewcommand{\thesection}{S\arabic{section}} 
\setcounter{section}{0}
\renewcommand{\thefigure}{S\arabic{figure}}
\setcounter{figure}{0}
\renewcommand{\theequation}{S\arabic{equation}}
\setcounter{equation}{0}
\FloatBarrier

{\huge Supplemental Materials}

\section{Materials and Methods}

\subsection{Calculating \VE from a Bernoulli infection model}
To clearly demonstrate the drivers of time-varying \VE estimates using cumulative attack rate ratios (ARR), we consider a simplified model of the infection process. In this model, each exposure event acts as a Bernoulli trial, leading to infection with probability $y$. Without vaccination, the cumulative probability of infection after $n$ exposure events is:
\begin{equation}
    P_U(n) = 1 - (1 - y)^n. 
\end{equation}
This can be thought of as the cumulative infection probability for a typical unvaccinated individual.

However, vaccinated individuals benefit from additional protection. The model of vaccine-derived protection depends on a vaccine's mechanism of action. For an all-or-nothing vaccine, vaccinated individuals are fully protected with probability $\Phi$ and receive no protection with probability $1-\Phi$. This means the cumulative infection probability for a typical vaccinated person is given by
\begin{equation}
    P_\text{all-or-nothing}(n) = \Phi \cdot 0 + (1 - \Phi) [1 - (1 - y)^n]. 
\end{equation}
We define \VEmeasspace as the ratio of cumulative infection probabilities in the vaccinated versus unvaccinated community, which mirrors the expected value of ARR in a mixed-vaccination population:
\begin{equation}
    VE_\text{all-or-nothing}(n) = 1 - \frac{(1 - \Phi) [1 - (1 - y)^n]}{1 - (1 - y)^n} = \Phi.
\end{equation}
Thus, $\VEmeasmath=\Phi$ for all-or-nothing vaccines.

For a leaky vaccine, each exposure is less likely to lead to successful establishment of infection in a vaccinated individual. If $\alpha$ is the per-exposure infection risk reduction, then the cumulative risk of infection for a typical vaccinated person is:
\begin{equation}
P_\text{leaky}(n) = P_\text{leaky}(n) = 1 - [1 - y\,(1 - \alpha)]^n.
\end{equation}
Calculating \VEmeasspace as above gives
\begin{equation}
    VE_\text{leaky}(n) = 1 - \frac{1 - [1 - y\,(1 - \alpha)]^n}{1 - (1 - y)^n}.
\end{equation}
For a leaky vaccine, \VEmeasspace varies non-linearly with the number of exposures $n$ and per-exposure risk reduction. 

If a vaccine provides heterogeneous protection across recipients, we define $K$ classes of vaccine-derived protection. A vaccinated individual will receive a per-exposure risk reduction of $\alpha_i$ if they are in protection class $i$, with probability mass $w_i$. \VEmeasspace is then given by
\begin{equation}
    VE_\text{heterogeneous} = 1 - \frac{\sum_{i=1}^K w_i \left[1 - \left(1 - y \left(1 - \alpha_i\right)\right)^n\right]}{1 - (1 - y)^n}.
\end{equation}
Similar to the leaky model, \VEmeasspace varies with the number of exposures $n$.

\begin{figure}
    \centering
    \includegraphics[width=0.8\linewidth]{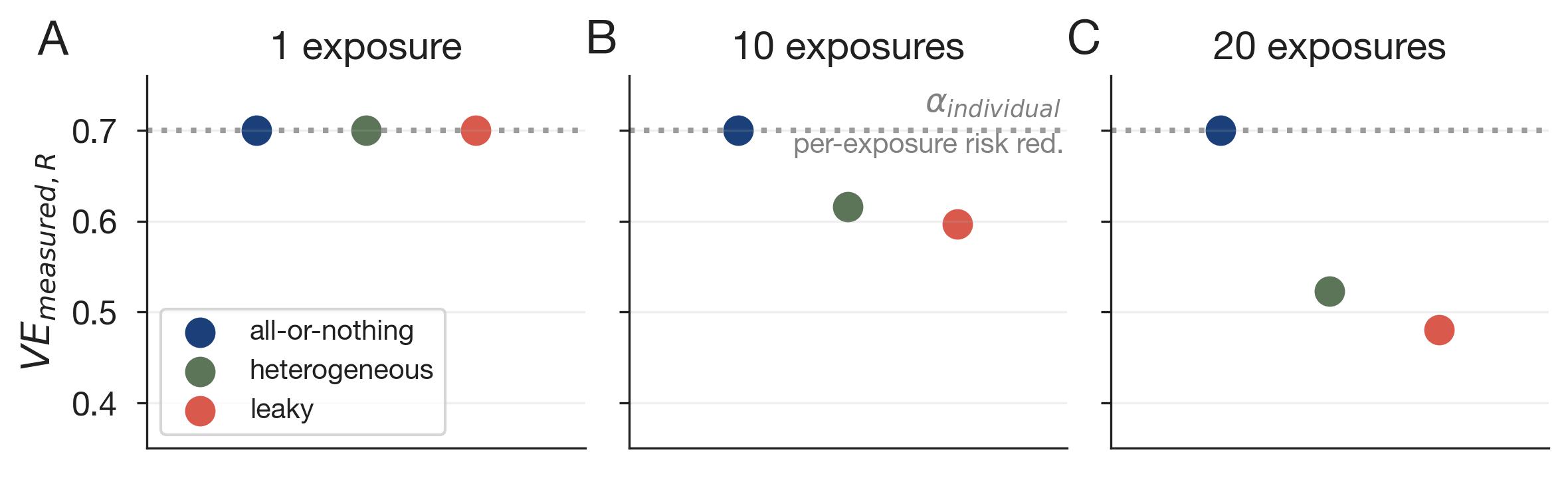}
    \caption{{\bf The difference between per-exposure (\VEind) and population-level vaccine effectiveness (\VEmeas) depends on mechanism of vaccine action and number of exposures.} \VEmeasspace is calculated using cumulative attack rates from the Bernoulli infection model when per-exposure $\VEindmath=0.7$ after 1 (A), 10 (B), and 20 (C) exposure events, with a per-exposure infection probability of 10\%. }
    \label{sfig:scatter_VE_bias}
\end{figure}

In order to compare \VEmeasspace across vaccine mechanisms, we define \VEindspace as the expected per-exposure risk reduction for a typical vaccinated individual, i.e., $\VEindmath = \Phi,\  \bar{\alpha},$ and $ \alpha$ for all-or-nothing, heterogeneous, and leaky vaccines, respectively. Comparing the three vaccine mechanisms when $\VEindmath=0.7$ and only a single exposure has occurred for each individual in the study, we observe $\VEmeasmath=0.7$ regardless of the vaccine's mechanism of action~(Fig.~\ref{sfig:scatter_VE_bias}A). Empirically, this scenario parallels a vaccine challenge study. For an all-or-nothing vaccine, subsequent exposures do not impact \VEmeas. However, subsequent exposures lead to reduced $\VEmeasmath < \VEindmath$ for both leaky and heterogeneous vaccines, and the difference grows as the number of exposures increases (Fig.~\ref{sfig:scatter_VE_bias}B,C). A risk-based \VE estimate for a leaky vaccine will be the same after a single exposure (i.e., $\VEmeasmath = 0.7$) , but will decline with subsequent exposures --- for example, to $\VEmeasmath = 0.597$ and $0.481$ after 10 and 20 exposures, respectively. For heterogeneous vaccines, the decline in risk-based \VE estimates with subsequent exposures is less.

\begin{figure}
    \centering
\includegraphics[width=0.8\linewidth]{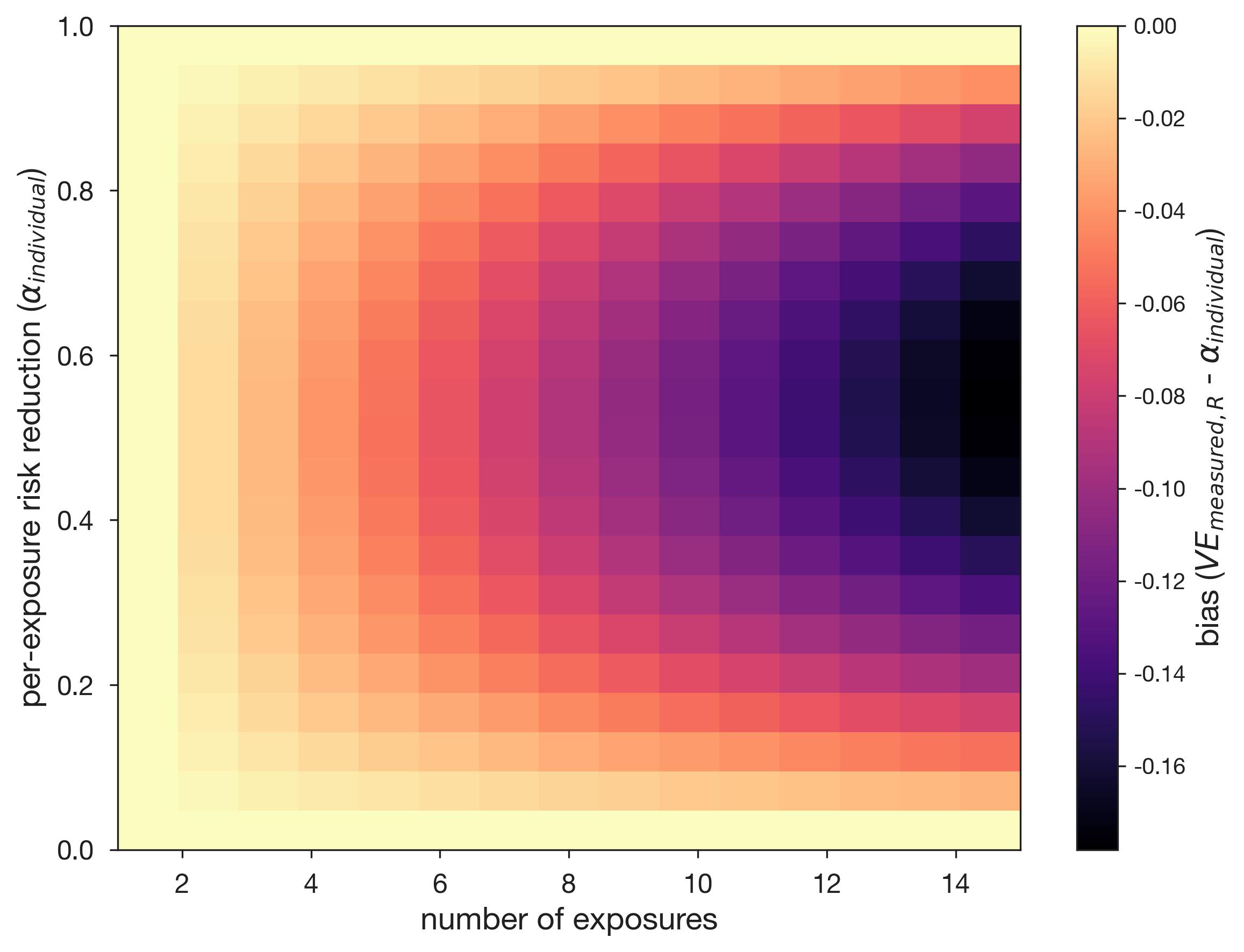}
    \caption{{\bf Bias between \VE and per-exposure risk reduction ($\alpha$) depends on the value of $\alpha$ and the number of exposures $n$.} Scenarios use the Bernoulli model with a 10\% baseline probability of infection per exposure, assuming a leaky vaccine.}
    \label{sfig:heatmap}
\end{figure}

The difference between \VEindspace and \VEmeasspace depends on both the number of exposure events and the value of \VEindspace (Fig.~\ref{sfig:heatmap}). Practically, this difference is likely to be higher when the basic reproduction number ($R_0$) is high and population-level immunity is low (i.e., there is little effect from indirect protection). We also note that HR estimators have been shown to increase with time for all-or-nothing vaccines~\cite{SMITH1984}, while their behavior depends on the distribution of protection for heterogeneous vaccines~\cite{Nikas2023Competing}.

\FloatBarrier

\begin{figure}
    \centering
\includegraphics[width=0.8\linewidth]{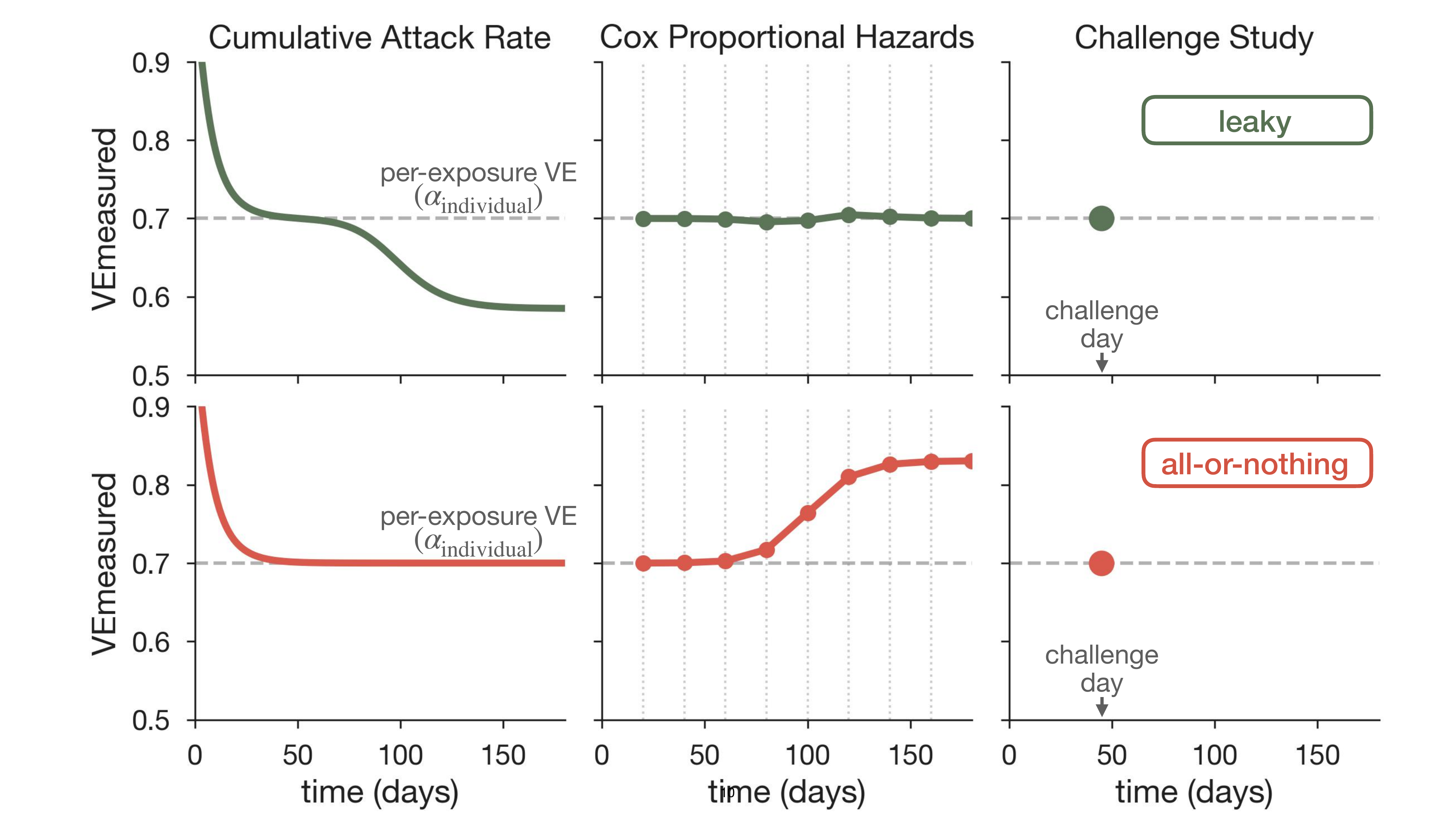}
    \caption{{\bf Estimates of vaccine effectiveness may vary through time and by vaccine mechanism depending on the study design and statistical method used.} Vaccine effectiveness (\VEmeas) estimates for leaky and all-or-nothing vaccines using cumulative attack rates, Cox proportional hazards with time-varying interactions, and challenge study risk.}
    \label{sfig:study_design}
\end{figure}

\subsection{Compartmental model formulation}

We simulate epidemic dynamics using a susceptible--infected--recovered (SIR) framework extended to include vaccination. In all model variants, the total population is $N$, a fraction $f$ are vaccinated at time $t = 0$, and the model's vaccine effectiveness parameter is given by $\mathrm{VE} \in [0, 1]$. The force of infection is
\begin{equation}
    \lambda(t) = \frac{\beta \bigl(I(t) + V_I(t)\bigr)}{N},
    \label{eq:foi}
\end{equation}
where $\beta$ is the per-contact transmission hazard, $I$ is the number of
infected unvaccinated individuals, and $V_I$ represents infected individuals
from the vaccinated population (defined differently for each vaccine mechanism
below). The recovery rate is $\gamma$, giving a mean infectious duration of
$\tau = 1/\gamma$ days. The basic reproduction number is $\mathcal{R}_0 = \beta/\gamma$.

All ordinary differential equations (ODEs) were solved using the LSODA method (Livermore Solver for Ordinary Differential Equations with Automatic stiffness detection) as implemented in \\ \texttt{scipy.integrate.solve\_ivp} (SciPy v1.7.3), with relative and absolute tolerances of $10^{-9}$ and $10^{-12}$ respectively and a maximum step size of 1 day. All simulations were conducted in Python 3.9.7.

\subsubsection{Leaky Vaccination}

Under a leaky mechanism, all vaccinated individuals experience an identical reduction in susceptibility ($\alpha$). The population is divided
into unvaccinated compartments $\{S, I, R\}$ and vaccinated compartments $\{
V_S, V_I, V_R \}$, with initial conditions
\begin{equation}
    S(0) = (1-f)\,S_0, \qquad V_S(0) = f\,S_0,
\end{equation}
where $S_0 \approx 1$ (fraction susceptible at $t=0$) and $f$ is the vaccine coverage at model initiation. The dynamics are
\begin{align}
    \frac{dS}{dt}   &= -\lambda S, \\
    \frac{dI}{dt}   &= \lambda S - \gamma I, \\
    \frac{dR}{dt}   &= \gamma I, \\[4pt]
    \frac{dV_S}{dt} &= -(1 - \alpha)\,\lambda\, V_S, \\
    \frac{dV_I}{dt} &= (1 - \alpha)\,\lambda\, V_S - \gamma V_I, \\
    \frac{dV_R}{dt} &= \gamma V_I,
    \label{eq:leaky}
\end{align}
where $\alpha$ is informed using an empirical \VE estimate.

\subsubsection{All-or-Nothing Vaccination}

Under an all-or-nothing mechanism, a fraction $\Phi$ of vaccinated
individuals are fully protected and placed in a static compartment $V_\emptyset$, while
the remaining fraction $(1 - \Phi)$ are completely unprotected and
placed in $V_S$. Initial conditions are
\begin{equation}
    S(0)   = (1-f)\,S_0, \qquad
    V_S(0) = (1-\Phi)\,f\,S_0, \qquad
    V_\emptyset(0)   = \Phi \cdot f\,S_0.
\end{equation}
The dynamics of the unvaccinated compartments are identical to
the leaky model, while the vaccinated compartments are defined by
\begin{align}
    \frac{dV_S}{dt} &= -\lambda\, V_S, \\
    \frac{dV_I}{dt} &= \lambda\, V_S - \gamma V_I, \\
    \frac{dV_R}{dt} &= \gamma V_I, \\
    \frac{dV_\emptyset}{dt}   &= 0.
\end{align}
Note that $V_S$ individuals are
fully susceptible ($\alpha = 0$), so this mechanism produces exactly the
same dynamics as the leaky model in the limit of low vaccination coverage~\cite{Halloran1992Interpretation}.

\subsubsection{Heterogeneous Vaccination}

To capture individual-level heterogeneity in vaccine response, we extend the
leaky model to $K$ susceptibility strata. Each stratum $k \in \{1, \ldots,
K\}$ contains a fraction $p_k$ of vaccinated individuals (where $\sum_k p_k
= 1$) and has a class-specific $\alpha$ of $\alpha_k$. The vaccinated population is
thus divided into $3K$ compartments $\{V_{S,k},\, V_{I,k},\, V_{R,k}\}$, and
the unvaccinated compartments $\{S, I, R\}$ follow the same equations as
before. The force of infection is
\begin{equation}
    \lambda(t) = \frac{\beta \Bigl(I(t) + \sum_{k=1}^{K} V_{I,k}(t)\Bigr)}{N},
\end{equation}
and the dynamics within each vaccinated stratum are
\begin{align}
    \frac{dV_{S,k}}{dt} &= -(1 - \alpha_k)\,\lambda\, V_{S,k}, \\
    \frac{dV_{I,k}}{dt} &= (1 - \alpha_k)\,\lambda\, V_{S,k} - \gamma V_{I,k}, \\
    \frac{dV_{R,k}}{dt} &= \gamma V_{I,k}.
    \label{eq:frailty}
\end{align}
Initial conditions for stratum $k$ are $V_{S,k}(0) = p_k \cdot f \cdot S_0$, where $f$ and $S_0$ are defined identically to the leaky vaccine.

\paragraph{Discretizing the Beta distribution.}
The stratum \VE levels $\alpha_k$ and mixing weights $p_k$ are derived by
discretizing a Beta distribution with mean $\bar{\alpha}$ and shape parameter
$c$~\cite{Gomes2014Missing,Nikas2023Competing}. Specifically, we set
\begin{equation}
    a = \bar{\alpha} c, \qquad b = (1 - \bar{\alpha})c,
\end{equation}
so that the mean vaccine effect parameter is $\bar{\alpha} = a/(a+b)$ and the variance is $\bar{\alpha}(1-\bar{\alpha})/(c+1)$. A larger $c$
yields a more concentrated distribution. We partition $[0, 1]$ into $K$
bins. The first and last bins span $[0,\,\delta]$ and $[1-\delta,\,1]$
respectively, where $\delta = \min(0.025,\, 1/K)$, and the remaining $K-2$
interior bins are equally spaced over $[\delta,\, 1-\delta]$. The midpoints of
the first and last bins are pinned to $\alpha_1 = 0$ and $\alpha_K = 1$ to
ensure the support of the distribution includes complete non-response and
complete protection. Interior midpoints are taken as the arithmetic mean of
their bin edges. 

The mixing weight for bin $k$ is
\begin{equation}
    p_k = F(e_{k+1};\, a, b) - F(e_k;\, a, b),
\end{equation}
where $e_k$ and $e_{k+1}$ are the left and right edges of bin $k$ and $F(\,
\cdot\,;\, a, b)$ is the Beta CDF, with the vector $(p_1, \ldots, p_K)$
re-normalized to sum to one to correct for numerical rounding. For
$c > 1{,}000$ the Beta CDF is replaced by its normal approximation
$\mathcal{N}(\bar{\alpha},\, \sigma^2)$ with $\sigma^2 = \bar{\alpha}(1-\bar{\alpha})/(c+1)$ to avoid
precision failures in the incomplete-beta function; for $c < 1$ the
distribution is approximated as Bernoulli with $p_1 = 1 - \bar{\alpha}$ and $p_K =
\bar{\alpha}$ (all remaining weights zero).

\subsubsection{Modeled Vaccine Effectiveness}

In all model variants, the observed VE in the modeled population (\VEmod) is estimated from the
simulated cumulative attack rates at the end of the epidemic ($t = T$):
\begin{equation}
    \VEmodmath = 1 - \frac{\Omega_v}{\Omega_u},
    \qquad
    \Omega_u = \frac{R(T)}{N_u}, \qquad
    \Omega_v = \frac{\sum_k V_{R,k}(T)}{N_v},
    \label{eq:VE_obs}
\end{equation}
where $N_u = (1-f)N$ and $N_v = fN$ are the initial unvaccinated and
vaccinated population sizes, respectively. This estimator corresponds to the
cohort attack-rate ratio used in prospective efficacy trials~\cite{Halloran1997Study}.

\subsection{Adjusted Model Parameterization}

A key challenge in using empirically measured vaccine effectiveness to
parameterize a leaky or heterogeneous SIRV model is that the model's
\emph{observed} VE---computed from simulated cumulative attack rates at
epidemic end (\eqref{eq:VE_obs})---will generally differ from the
input susceptibility-reduction parameter $\alpha$ (leaky) or the mean of
the Beta distribution $\bar{\alpha}$ (heterogeneous) due to differential depletion of susceptibles. We therefore solve for an adjusted
parameter value that makes the model-predicted observed \VE (\VEmod) match the
empirically measured target \VEmeas exactly.

\subsubsection{Leaky Model Adjusted Parameterization}

For the leaky model, the adjusted susceptibility-reduction parameter
$\alpha^*$ is defined as the value that satisfies
\begin{equation}
    g(\alpha^*) \;=\; \VEmeasmath,
    \label{eq:leaky_adj}
\end{equation}
where $g(\alpha) = 1 - \Omega_v(\alpha)/\Omega_u(\alpha)$ and
$\Omega_u(\alpha)$, $\Omega_v(\alpha)$ are the unvaccinated and vaccinated
cumulative attack rates obtained from the leaky model final-size equations
(\eqref{eq:final_size}). We solve~\eqref{eq:leaky_adj} using bisection
(\texttt{scipy.optimize.root\_scalar}, method \texttt{bisect}) over the
bracket $[\alpha_\ell,\, \alpha_u]$, where
\begin{equation}
    \alpha_\ell = \max\,\bigl(0,\; \VEmeasmath \bigr),
    \qquad
    \alpha_u = \min\!\left(1,\; \frac{1 - \mathcal{R}_0^{-1}}{f}\right).
\end{equation}
The lower bound limits the potential solution space to $\alpha^* \geq \VEmeasmath$, given that $\VEmeasmath \leq \alpha$ for any $\alpha$ under the leaky model.
The upper bound is the \VE threshold at which herd immunity would be achieved for vaccination coverage $f$. Values of $\alpha$ above
this threshold prevent epidemic spread entirely and thus do not contribute to progress for our solver. If $\alpha_u \leq \VEmeasmath$ (i.e., herd immunity
effects preclude reaching the target \VE), $\alpha^*$ is set directly to
\VEmeasspace without adjustment.

To initialize the bisection, we supply an initial guess for the attack rates
derived from the effective reproduction number $\mathcal{R}_e = \mathcal{R}_0
(1 - f\cdot \alpha)$. Letting $\tilde{Z} = 1 - \exp\!\bigl[-\tilde{a}(1 -
\mathcal{R}_e^{-1})\bigr]$ with scaling parameter $\tilde{a} = (1-f)(1 -
\alpha)\,\mathcal{R}_0 \times 50$, the initial guesses are
\begin{equation}
    \tilde{\Omega}_u = 1 - e^{-\mathcal{R}_0 \tilde{Z}},
    \qquad
    \tilde{\Omega}_v = 1 - e^{-(1-\alpha)\,\mathcal{R}_0 \tilde{Z}}.
\end{equation}

\subsubsection{Heterogeneous Model Adjusted Parameterization}

For the heterogeneous model, the concentration parameter $c$ is held fixed, and the adjusted mean \VE parameter $\bar{\alpha}^*$ is defined as the value satisfying
\begin{equation}
    h(\bar{\alpha}^*) \;=\; \VEmeasmath,
    \label{eq:frailty_adj}
\end{equation}
where $h(\bar{\alpha}) = 1 - \bar{\Omega}_v(\bar{\alpha}) / \Omega_u(\bar{\alpha})$,
$\Omega_u(\bar{\alpha})$ and $\{\Omega_{v,k}(\bar{\alpha})\}$ are the unvaccinated and
per-stratum vaccinated cumulative attack rates from the heterogeneous model
final-size equations (\eqref{eq:final_size_frailty}), and
\begin{equation}
    \bar{\Omega}_v(\bar{\alpha})
    = \sum_{k=1}^{K} \frac{p_k(\bar{\alpha})}{f} \,\Omega_{v,k}(\bar{\alpha})
\end{equation}
is the population-weighted average vaccinated attack rate. Here $p_k(\bar{\alpha})$
denotes the stratum mixing weights generated by discretizing
$\mathrm{Beta}(a,b)$ with $a = \bar{\alpha} c$ and $b = (1-\bar{\alpha})c$.
\eqref{eq:frailty_adj} is solved by bisection over the bracket
$[\bar{\alpha}_\ell,\, \bar{\alpha}_u]$, where
\begin{equation}
    \bar{\alpha}_\ell = \max\!\bigl(0.01,\; \VEmeasmath\bigr),
    \qquad
    \bar{\alpha}_u = 0.99,
\end{equation}
with the lower bound again enforcing $\bar{\alpha}^* \geq \VEmeasmath$.
If no root exists in $[\bar{\alpha}_\ell, \bar{\alpha}_u]$---which can occur when
$\VEmeasmath \geq 0.99$---the endpoint yielding VE closest to
the target is returned. The final-size system for the heterogeneous model
is itself solved numerically at each bisection iteration using\\
\texttt{scipy.optimize.fsolve}, initialized with per-stratum attack rate
guesses derived from the effective reproduction number $\mathcal{R}_e =
\mathcal{R}_0(1 - f\,\bar{\alpha})$, where $\bar{\alpha} = \sum_k p_k
\alpha_k$ is the population-weighted mean susceptibility reduction. Letting
$\tilde{a} = (1 - f)(1 - \bar{\alpha})\,\mathcal{R}_0 \times 50$, the
initial guesses are
\begin{align}
    \tilde{\Omega}_u &= 1 - e^{-\mathcal{R}_0 \tilde{Z}} \\
    \tilde{\Omega}_{v,k} &= 1 - e^{-(1-\alpha_k)\,\mathcal{R}_0 \tilde{Z}}, \quad k = 1,\ldots,K
\end{align}
with $\tilde{Z} = 1 - \exp\!\bigl[-\tilde{a}(1 - \mathcal{R}_e^{-1})\bigr]$.

\newpage
\FloatBarrier

\begin{figure}
    \centering
\includegraphics[width=0.8\linewidth]{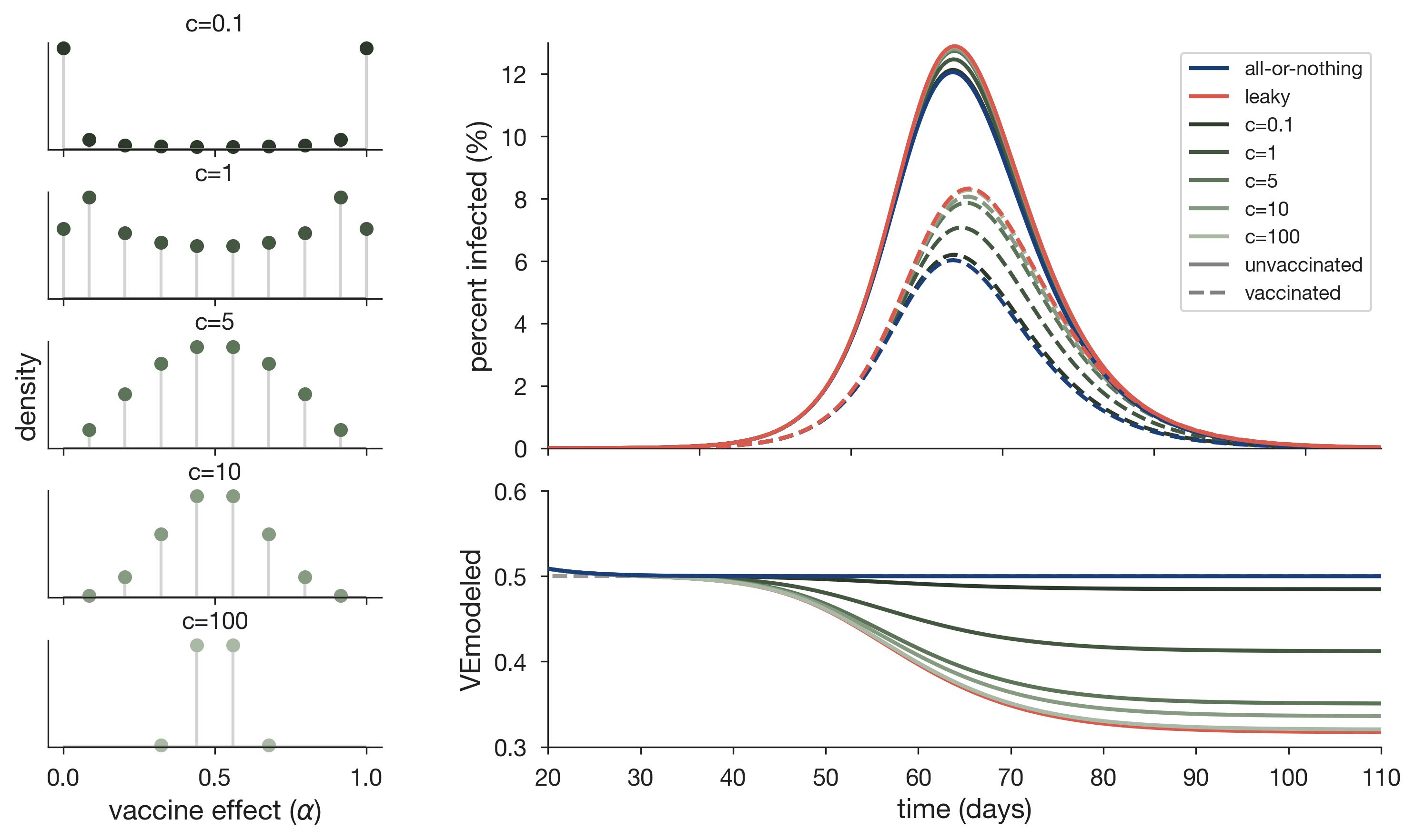}
    \caption{{\bf Heterogeneous model infection dynamics vary based distribution of vaccine-derived protection.} Temporal infection dynamics for vaccinated (dashed) and unvaccinated (solid) populations, and time-varying \VEmodspace for all-or-nothing, leaky, and multiple heterogeneous protection shape parameters ($c$, as labeled) are depicted under the same parameterization ($\alpha=\Phi=\bar{\alpha}=0.5$, $R_0=2.2$, 35\% vaccine coverage).}
    \label{sfig:time-varying-VE-supp}
\end{figure}

\begin{figure}
    \centering
    \includegraphics[width=0.8\linewidth]{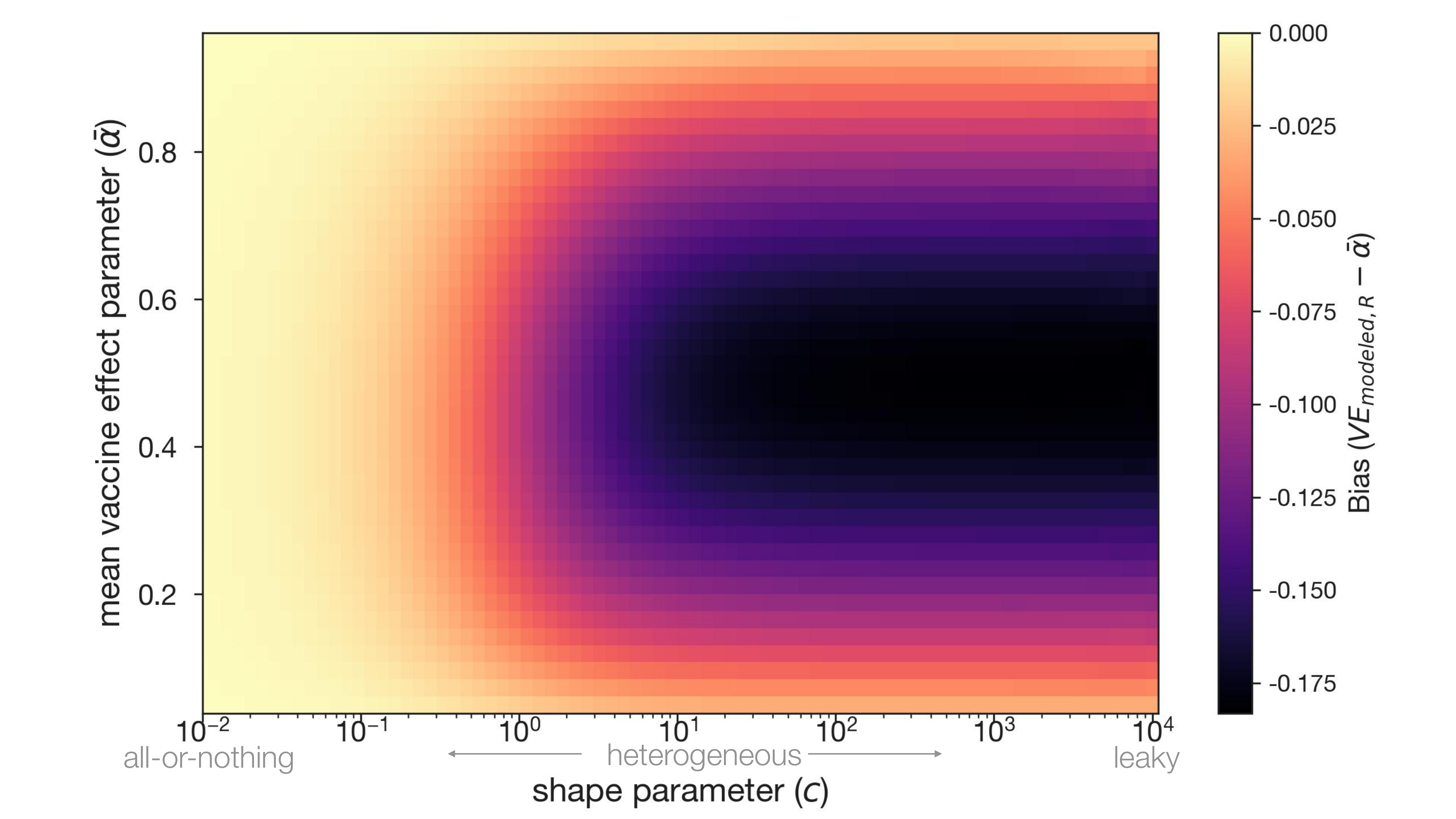}
    \caption{{\bf The magnitude of heterogeneous model bias depends on both the mean vaccine effect and the distribution of vaccine-derived protection.} Bias is calculated as the difference between \VEmodspace and $\bar{\alpha}$ when $R_0=2.2$, $\gamma=1/4$, $v=0.35$, and $K=50$.}
    \label{sfig:frailty_heatmap}
\end{figure}

\begin{figure}
    \centering
    \includegraphics[width=0.8\linewidth]{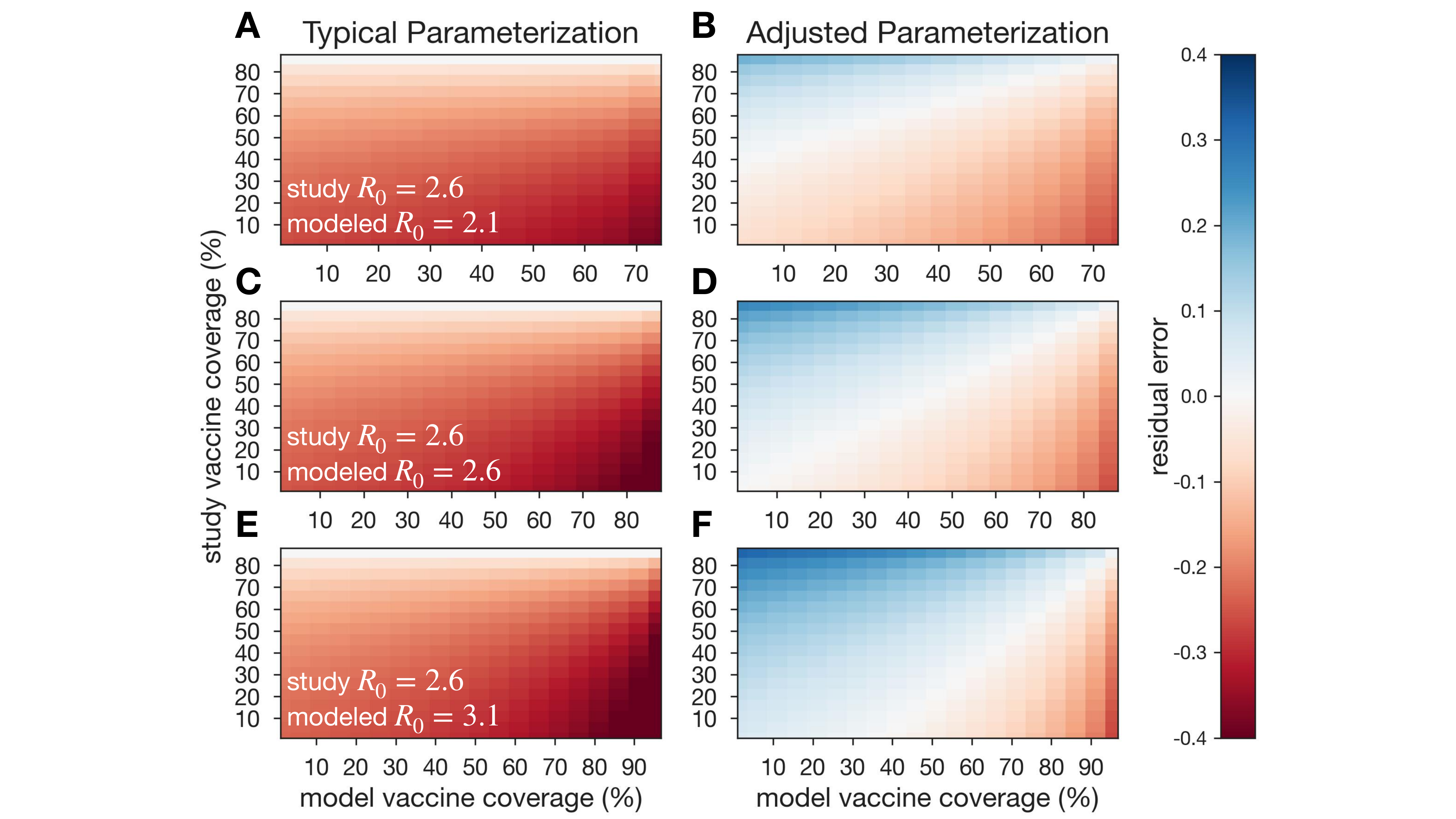}
    \caption{{\bf Adjusted parameterization accuracy for leaky model depends on the agreement of vaccine coverage and $R_0$ values between model and study.} \VEmeasspace is estimated from a study scenario with $R_0 = 2.6$, $\gamma=1/4$, and vaccine provides a hazard reduction of $\alpha=0.7$. \VEmodspace is estimated with and without adjusted parameterization using the vaccine coverage and $R_0$ shown.}
    \label{sfig:correction_accuracy}
\end{figure}

\begin{figure}
    \centering
    \includegraphics[width=0.8\linewidth]{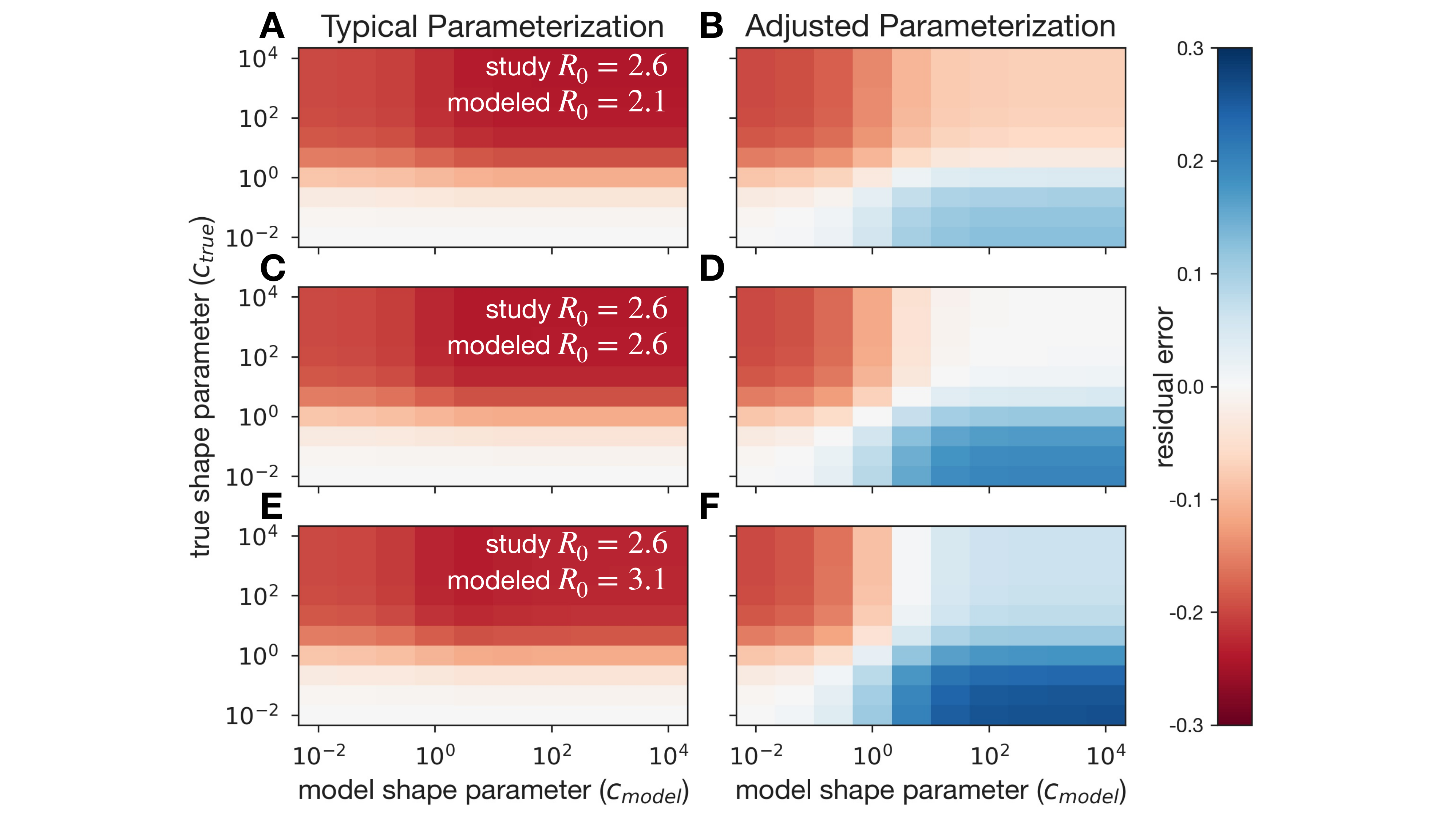}
    \caption{{\bf Adjusted parameterization accuracy for heterogeneous model depends on the agreement of the vaccine-derived protection distribution shape ($c$) and $R_0$ values between model and study.} \VEmeasspace is estimated from a study scenario with $R_0 = 2.6$, $\gamma=1/4$, $K=50$, and vaccine provides a hazard reduction of $\bar{\alpha}=0.7$. \VEmodspace is estimated with and without adjusted parameterization  using the shape parameter and $R_0$ shown.}
    \label{sfig:heterogeneous_correction_heatmaps}
\end{figure}

\begin{figure}[h!]
    \centering
\includegraphics[width=0.8\linewidth]{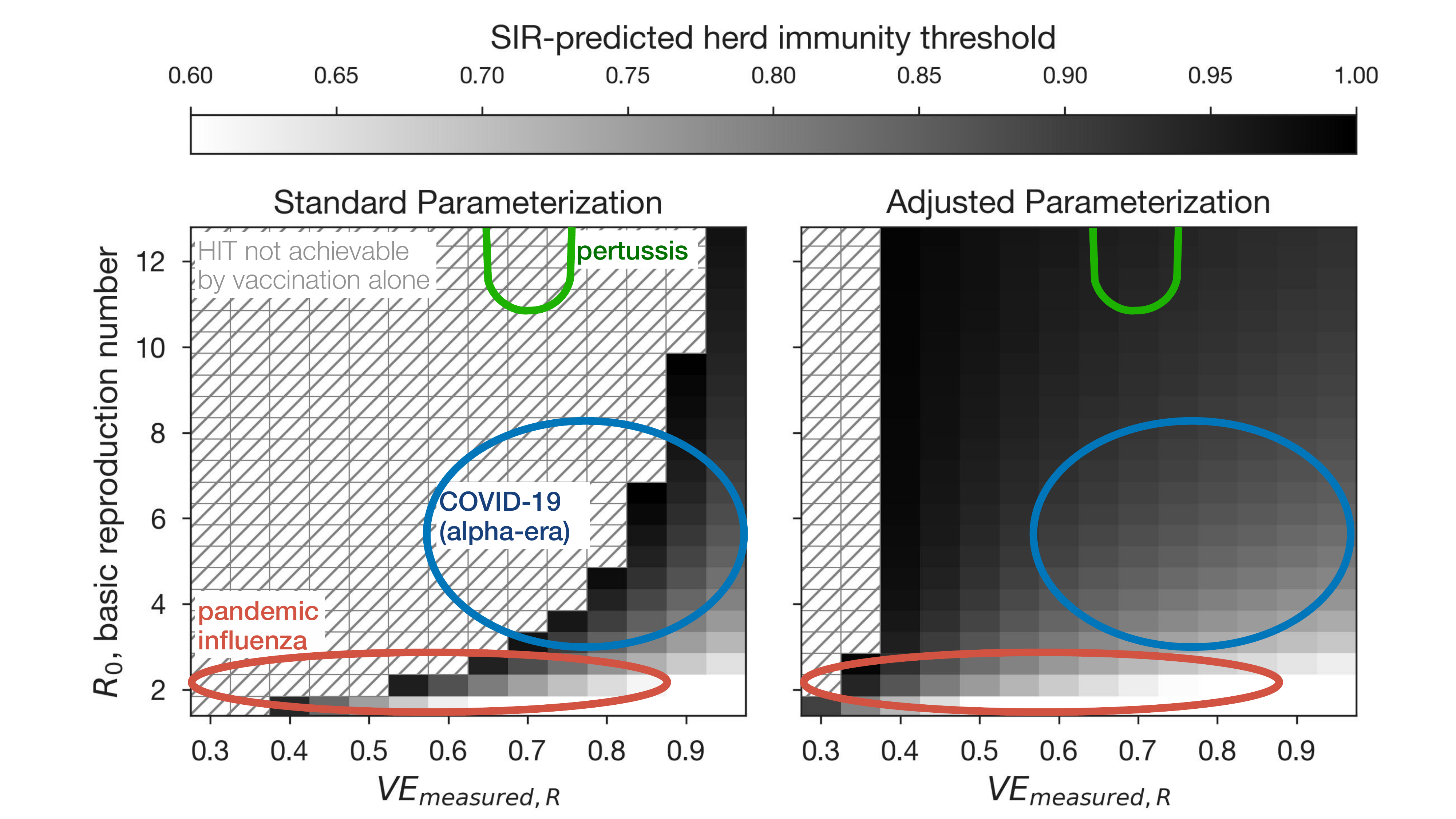}
    \caption{{\bf Adjusted parameterization of leaky model leads to lower predicted herd immunity thresholds.} Plausible parameter ranges are colored for pandemic influenza, COVID-19, and pertussis.}
    \label{sfig:HIT_heatmaps}
\end{figure}

\end{document}